\documentclass{article}
\usepackage{amssymb}
\usepackage{amsmath}
\usepackage{subfigure}
\newtheorem{rem}{Remark}[section]
\newcommand{\br}{\begin{rem}}
\newcommand{\er}{\end{rem}}
\newtheorem{ex}[rem]{Example}
\newcommand{\bex}{\begin{ex}}
\newcommand{\eex}{\end{ex}}
\newtheorem{Def}[rem]{Definition}
\newcommand{\bd}{\begin{Def}}
\newcommand{\ed}{\end{Def}}
\newtheorem{theorem}[rem]{Theorem}
\newcommand{\bt}{\begin{theorem}}
\newcommand{\et}{\end{theorem}}
\newtheorem{prop}[rem]{Proposition}
\newcommand{\bp}{\begin{prop}}
\newcommand{\ep}{\end{prop}}
\newtheorem{lemma}[rem]{Lemma}
\newcommand{\bl}{\begin{lemma}}
\newcommand{\el}{\end{lemma}}
\newtheorem{corollary}[rem]{Corollary}
\newcommand{\bc}{\begin{corollary}}
\newcommand{\ec}{\end{corollary}}
\newcommand{\be}{\begin{equation}}
\newcommand{\ee}{\end{equation}}
\newcommand{\bea}{\begin{eqnarray}}
\newcommand{\eea}{\end{eqnarray}}
\newcommand{\pa}{\partial}
\newcommand{\nn}{\nonumber}
\newcommand{\adots}{\mathinner{\mkern2mu\raise1pt\hbox{.}\mkern2mu
\raise4pt\hbox{.}\mkern2mu\raise7pt\hbox{.}\mkern1mu}}

\title{Recursive Procedures for Krall-Sheffer Operators}
\author{Allan P. Fordy and Michael J. Scott,
  \\ School of Mathematics, \\
University of Leeds. \\ Leeds LS2 9JT, UK.\\
e-mail: A.P.Fordy@leeds.ac.uk}

\begin{document}

\maketitle

\begin{abstract}
We consider the Krall-Sheffer class of admissible, partial differential
operators in the plane.  We concentrate on algebraic structures, such as the role of
commuting operators and symmetries.  For the polynomial eigenfunctions, we give explicit forms of the
the $3-$level recurrence relations and differential raising operators, which are shown to satisfy unusual commutation relations.
We present new generating functions for two of the cases.
\end{abstract}


\section{Introduction}

Orthogonal polynomials appear in many places in mathematics, physics and engineering. In mathematical physics, orthogonal
polynomials arise in the context of eigenfunctions of linear, second order differential operators.
Bochner (see \cite{11-1}, p150) classified those operators which have exactly one polynomial eigenfunction
of each degree.
The classical, one-dimensional examples possess important structures which enable their explicit
construction.  Among these are the $3-$point recurrence relations and the differential ladder
operators, which allow the polynomials to be built in a recursive manner.  Starting from Rodrigues
formulae and generating functions, these structures can themselves be built in a systematic
way.

The classification of orthogonal polynomials in the plane and higher
dimensional spaces is much more complicated than the case of the real line.  However, requiring the
polynomials to be eigenfunctions of linear differential operators makes the classification problem more feasible.
An early and systematic treatment of such polynomials in two variables, satisfying a {\em second order} eigenvalue problem, was
given by Krall and Sheffer \cite{67-2}.  They showed that, up to affine transformation,
there are 9 distinct cases.  Three of these just correspond to products of classical $1-$dimensional polynomials,
but the remaining six are non-trivial extensions.

To generalise Bochner's result for the classical orthogonal polynomials, Krall and Sheffer \cite{67-2} introduced the concept of an admissible operator.  In two dimensions we have to expect the existence of $N+1$ polynomial eigenfunctions of degree $N$.

\bd[Admissible Partial Differential Operator]\label{def:admiss}
The differential operator $L$ is said to be {\em admissible} if, for the eigenvalue problem
$$
L \varphi = \lambda \varphi,
$$
there exists an infinite sequence of eigenvalues $\{\lambda_N\}_{N=0}^\infty$, such that for each
$\lambda=\lambda_N$ there are no non-zero polynomial solutions of degree $d<N$
and there are exactly $N+1$ linearly independent polynomial solutions of degree $N$.
\ed
It is easily seen that a {\em second order} admissible differential operator must have the form
\be                               \label{admiss}
\begin{array}{l}
L \varphi:= (\alpha x^2 + d_1 x + e_1 y+f_1) \varphi_{xx} + (2\alpha x y + d_2 x + e_2 y+f_2) \varphi_{xy}\\[2mm]
 \hspace{2cm} + (\alpha y^2 +d_3 x + e_3 y+f_3) \varphi_{yy} + (\beta x +\kappa_1) \varphi_x + (\beta y + \kappa_2) \varphi_y.
 \end{array}
\ee
The number of parameters is reduced further by making affine transformations on the $x-y$ space, giving $9$ canonical forms (see Section \ref{geom}).
The degree $N$ polynomial eigenfunctions form an $N+1$ dimensional vector space $V_N$ and we may choose a basis of {\em monic} polynomials:
\be  \label{pmn}  %
P_{m,n} = x^m y^n + \mbox{lower order terms} ,\quad\mbox{for}\;\;\; m+n=N .
\ee  %
The eigenvalue for $P_{m,n}$ is easily calculated from the leading order term to be
\be\label{lambda}  %
\lambda_{m+n}=(m+n) ((m+n-1)\alpha+\beta).
\ee  %
We see that when $\beta$ is not a negative integer, $\lambda_{m_1+n_1}\neq\lambda_{m_2+n_2}$ for $m_1+n_1\neq m_2+n_2$.

This family of polynomial eigenfunctions can be arranged into a triangular array (see Figure \ref{trilatt}), with each horizontal level representing a particular $V_N$.  In place of the {\em three point recurrence relations} for the one dimensional case,  Krall and Sheffer showed that in two-dimensions there are two {\em three {\bf level} recurrence relations} of the form
\bea  %
P_{m+1,n} &=& x P_{m,n} +\sum_{i=0}^{m+n} a_{i,m,n} P_{i,m+n-i}+\sum_{i=0}^{m+n-1}b_{i,m,n} P_{i,m+n-1-i}, \nn\\[0mm]
&&  \label{ks3level}  \\[0mm]
P_{m,n+1} &=& y P_{m,n} +\sum_{i=0}^{m+n} c_{i,m,n} P_{m+n-i,i}+\sum_{i=0}^{m+n-1}d_{i,m,n} P_{m+n-1-i,i}, \nn
\eea  %
but gave no specific forms, apart from the case of direct products of classical one dimensional polynomials.
Such 3 level recurrence relations have been discussed more generally in \cite{95-6,01-11,06-4}.  A detailed discussion of orthogonal polynomials in two variables can be found in \cite{99-13}.

The coefficients of the second order terms of the operator (\ref{admiss}) can be considered as the components of an upper index metric tensor, with $L$ being a deformation of the corresponding Laplace-Beltrami operator $L_b$.  In \cite{03-4,01-1} it was observed that each of the 9 metrics was either {\em flat} or {\em constant curvature} and that the additional first order terms correspond to ``trivial'' magnetic field, which can therefore be gauged to a purely potential term.  They also showed that each of these cases (considered as a quantum system) was {\em super-integrable} and presented two commuting operators for each case.

In this paper we first discuss the symmetries of these Laplace-Beltrami operators and use these to group together the 9 canonical forms into 4 classes (see Section \ref{geom}).  Within each class the metrics are related through (non-affine) coordinate transformations.  Each metric is invariant with respect to an involutive transformation, corresponding to an involutive automorphism of the algebra of symmetries. These involutions are very useful in Section \ref{super}, where we discuss super-integrability.  We briefly review separable potentials, defining $L=L_b+U(x,y)$, each of which is characterised by the existence of a commuting, second order operator and each of which depends upon two arbitrary functions of one variable (as in the classical mechanical case; see \cite{76-8}).  Requiring two such commuting operators (for super-integrability) reduces these potentials to simple rational functions depending upon (at most) 3 arbitrary parameters.
Super-integrability (with linear or quadratic commuting operators) is associated with separability in more than one coordinate system (see, for example, \cite{99-11}). It is then possible to gauge transform these super-integrable cases to Krall-Sheffer form (just the reverse calculation of \cite{01-1}), but this sometimes restricts the parameters to a sub-case.

Our main results are in Sections \ref{pols} and \ref{genfuns}.  In the first of these, we discuss the algebraic structure of the polynomial eigenfunctions of the Krall-Sheffer operators.  We first emphasise the connection between {\em super-integrability} and the construction of eigenfunctions.  The commuting operators can be used to build the two-dimensional polynomial eigenfunctions out of one-dimensional polynomials, satisfying a reduced differential operator.  In Section \ref{pols}, we also give {\em explicit formulae} for the 3-level recurrence relations for all the (non-trivial) Krall-Shefffer operators.  We also present {\em explicit formulae} for a pair of differential ladder operators, which satisfy $R_{+,x}P_{m,n}=P_{m+1,n},\; R_{+,y}P_{m,n}=P_{m,n+1}$ ({\em raising operators}).  None of these formulae have previously appeared in the literature.

In Section \ref{genfuns} we present {\em generating functions} for the polynomials of Cases V and VIII (both new) and of Case IX.  The relationship (discussed in Section \ref{geom}) between Cases I and IX, gives us an array of eigenfunctions (not all polynomial) of a restricted form of the Case I operator.

\section{The Geometry of Krall-Sheffer Operators}\label{geom}

For an $n-$dimensional (pseudo-)Riemannian space, with local coordinates
$x^1,\cdots ,x^n$ and metric $g_{ij}$, the Laplace-Beltrami operator is defined
by $L_b f = g^{ij} \nabla_i\nabla_j f$, which has explicit form  %
\be                \label{lbf}   %
L_b f = \sum_{i,j=1}^n \frac{1}{\sqrt{g}}
\frac{\pa}{\pa x^j}\left(
               \sqrt{g} g^{ij}\frac{\pa f}{\pa x^i}\right),
\ee   %
where $g$ is the determinant of the matrix $g_{ij}$.  The coefficients of
leading order terms in the Laplace-Beltrami operator are the coefficients of
the {\em inverse metric} $g^{ij}$.  For a metric with isometries, the
infinitesimal generators (Killing vectors) are just first order differential
operators which commute with the Laplace-Beltrami operator (\ref{lbf}).  When
the space is either flat or constant curvature, it possesses the maximal group
of isometries, which is of dimension $\frac{1}{2}n(n+1)$.  In this case, $L_b$
is proportional to the second order {\em Casimir} function of the symmetry algebra
(see \cite{74-7}).  The Krall-Sheffer metrics fall into this category, with $n=2$.

When $n=2$ we have $3$ independent Killing vectors for flat and constant curvature metrics.  For the particular metrics defined by each of the Krall-Sheffer operators, a basis of Killing vectors is easy to calculate explicitly.  Four Lie algebras occur and the $9$ canonical forms can be grouped under these, with the members of each group being related through {\em nonlinear transformations} of the coordinates (see Table \ref{symtab}).
\begin{table}[hbt]
\begin{center}
\begin{tabular}{|c|c|}
  \hline
   Cases & Symmetry algebra \\ \hline\hline
  I, IX & $so(3)$ \\ \hline
  II, III & $sl(2)$ \\ \hline
  IV, VI, VII & $e(2)$ \\ \hline
  V, VIII & $e(1,1)$ \\
  \hline
\end{tabular}
\end{center}
\caption{The symmetry algebras for the Krall-Sheffer metrics} \label{symtab}
\end{table}
The algebras $so(3)$ and $sl(2)$ generate metrics with constant curvature, while the Euclidean and pseudo-Euclidean algebras generate flat metrics.
For the purposes of this grouping we label the coordinates $(x_i,y_i),\, i=1,\dots ,9$.

\subsection{The Rotation Algebra $so(3)$: Cases I and XI}\label{i-ix}

The abstract algebra, with ${\bf A}, {\bf B}$ and ${\bf C}$ satisfying
$$
[{\bf A},{\bf B}]={\bf C},\;\; [{\bf B},{\bf C}]={\bf A},\;\; [{\bf C},{\bf A}]={\bf B},\quad\mbox{has Casimir function}\quad
{\cal C} = {\bf A}^2+{\bf B}^2+{\bf C}^2.
$$
The concrete realisation
$$
{\bf A}=2 \sqrt{x_1(1-x_1-y_1)}\,\pa_{x_1},\quad {\bf B}=2 \sqrt{y_1(1-x_1-y_1)}\,\pa_{y_1},\quad {\bf C}=2 \sqrt{x_1y_1}\,(\pa_{x_1}-\pa_{y_1}),
$$
gives the Laplace-Beltrami operator
$$
L_b=-\frac{1}{4}{\cal C}=(x_1^2-x_1) \pa_{x_1}^2+2x_1y_1 \pa_{x_1}\pa_{y_1}+(y_1^2-y_1)\pa_{y_1}^2+\frac{1}{2}((3x_1-1)\pa_{x_1}+(3y_1-1)\pa_{y_1}).
$$
The concrete realisation
$$
{\bf A}=\sqrt{1-x_9^2-y_9^2}\,\pa_{x_9},\quad {\bf B}= \sqrt{1-x_9^2-y_9^2}\,\pa_{y_9},\quad {\bf C}=y_9\pa_{x_9}-x_9\pa_{y_9},
$$
gives the Laplace-Beltrami operator
$$
L_b=- {\cal C} =(x_9^2-1) \pa_{x_9}^2+2x_9y_9 \pa_{x_9}\pa_{y_9}+(y_9^2-1)\pa_{y_9}^2+2(x_9\pa_{x_9}+y_9\pa_{y_9}).
$$

In each case, the {\em involution}
$$
(\bar x,\bar y)=(y,x) \quad\mbox{gives the automorphism}\quad {\bf A}\leftrightarrow {\bf B},\;\; {\bf C}\rightarrow -{\bf C},
$$
which leaves the $L_b$ invariant.  The two realisations are related through the change of coordinates $x_1=x_9^2,\quad y_1=y_9^2$.

\subsection{The Special Linear Algebra $sl(2)$: Cases II and III}\label{ii-iii}

The abstract algebra, with ${\bf E}, {\bf F}$ and ${\bf H}$ satisfying
$$
[{\bf H},{\bf E}]=2{\bf E},\quad [{\bf H},{\bf F}]=-2{\bf F},\quad [{\bf E},{\bf F}]={\bf H},\quad\mbox{has Casimir function}\quad
{\cal C} = {\bf H}^2+2{\bf E}{\bf F}+2{\bf F}{\bf E}.
$$
The concrete realisation
$$
{\bf H}=4x_2\pa_{x_2},\quad {\bf E}=2 \sqrt{x_2y_2}\,\pa_{y_2},\quad {\bf F}=4 \sqrt{x_2y_2}\,\pa_{x_2}+2(y_2-1)\sqrt{\frac{y_2}{x_2}}\pa_{y_2},
$$
gives the Laplace-Beltrami operator
$$
L_b=\frac{1}{16} {\cal C} =x_2^2 \pa_{x_2}^2+2x_2y_2 \pa_{x_2}\pa_{y_2}+(y_2^2-y_2)\pa_{y_2}^2+\frac{1}{2}(3x_2\pa_{x_2}+(3y_2-1)\pa_{y_2}),
$$
which is invariant under the {\em involution}
$$
(\bar x_2,\bar y_2)=\left(\frac{(y_2-1)^2}{x_2},y_2\right), \quad\mbox{giving the automorphism}\quad
                                     {\bf E}\leftrightarrow {\bf F},\;\; {\bf H}\rightarrow -{\bf H}.
$$
The concrete realisation
$$
{\bf H}=4x_3\pa_{x_3}+2y_3\pa_{y_3},\quad {\bf E}= x_3\pa_{y_3},\quad {\bf F}=4y_3\pa_{x_3}+\frac{4x_3+3y_3^2}{x_3}\,\pa_{y_3},
$$
gives the Laplace-Beltrami operator
$$
L_b=\frac{1}{16} {\cal C} = x_3^2 \pa_{x_3}^2+2x_3y_3 \pa_{x_3}\pa_{y_3}+(y_3^2+x_3)\pa_{y_3}^2+\frac{3}{2}(x_3\pa_{x_3}+y_3\pa_{y_3}),
$$
which is invariant under the {\em involution}
$$
(\bar x_3,\bar y_3)=\left(\frac{(y_3^2+4x_3)^2}{x_3^3},\frac{(y_3^2+4x_3)y_3}{x_3^2}\right), \quad\mbox{giving the automorphism}\quad
                                     {\bf E}\leftrightarrow {\bf F},\;\; {\bf H}\rightarrow -{\bf H}.
$$
The two realisations are related through the change of coordinates $x_2=-64 x_3,\quad y_2=-y_3^2/(4x_3)$.

\subsection{The Euclidean Algebra $e(2)$: Cases IV, VI and VII}\label{iv-vii}

The abstract algebra, with ${\bf A}, {\bf B}$ and ${\bf C}$ satisfying
$$
[{\bf A},{\bf B}]=0,\quad [{\bf C},{\bf A}]={\bf B},\quad [{\bf B},{\bf C}]={\bf A},\quad\mbox{has Casimir function}\quad
{\cal C} = {\bf A}^2+{\bf B}^2.
$$
The concrete realisation
$$
{\bf A}=\sqrt{x_4}\pa_{x_4},\quad {\bf B}= \sqrt{y_4}\pa_{y_4},\quad {\bf C}=2\sqrt{x_4y_4}\,(\pa_{y_4}-\pa_{x_4}),
$$
gives the Laplace-Beltrami operator
$$
L_b= {\cal C} = x_4 \pa_{x_4}^2+y_4\pa_{y_4}^2+\frac{1}{2}(\pa_{x_4}+\pa_{y_4}),
$$
which is invariant under the {\em involution}
$$
(\bar x_4,\bar y_4)=(y_4,x_4), \quad\mbox{giving the automorphism}\quad {\bf A}\leftrightarrow {\bf B},\;\; {\bf C}\rightarrow -{\bf C}.
$$
The concrete realisation
$$
{\bf A}=\sqrt{x_6}\pa_{x_6},\quad {\bf B}= \pa_{y_6},\quad {\bf C}=\sqrt{x_6}\,(2\pa_{y_6}-y_6\pa_{x_6})
$$
gives the Laplace-Beltrami operator
$$
L_b= {\cal C} = x_6 \pa_{x_6}^2+\pa_{y_6}^2+\frac{1}{2}\pa_{x_6},
$$
which is invariant under the {\em involution}
$$
(\bar x_6,\bar y_6)=\left(\frac{y_6^2}{4},2\sqrt{x_6}\right), \quad\mbox{giving the automorphism}\quad {\bf A}\leftrightarrow {\bf B},\;\; {\bf C}\rightarrow -{\bf C}.
$$
The concrete realisation
$$
{\bf A}=\pa_{x_7},\quad {\bf B}=\pa_{y_7},\quad {\bf C}=x_7\pa_{y_7}-y_7\pa_{x_7}
$$
gives the Laplace-Beltrami operator
$$
L_b= {\cal C} = \pa_{x_7}^2+\pa_{y_7}^2,
$$
which is invariant under the {\em involution}
$$
(\bar x_7,\bar y_7)=(y_7,x_7), \quad\mbox{giving the automorphism}\quad {\bf A}\leftrightarrow {\bf B},\;\; {\bf C}\rightarrow -{\bf C}.
$$
These realisations are related through the changes of coordinates: $x_4=x_6,\; y_4=y_6^2$, and $x_6=x_7^2,\; y_6=y_7$.

\subsection{The Pseudo-Euclidean Algebra $e(1,1)$: Cases V and VIII}\label{v-viii}

The abstract algebra, with ${\bf E}, {\bf F}$ and ${\bf H}$ satisfying
$$
[{\bf H},{\bf E}]={\bf E},\quad [{\bf H},{\bf F}]=-{\bf F},\quad [{\bf E},{\bf F}]=0,\quad\mbox{has Casimir function}\quad
{\cal C} = {\bf E}{\bf F}.
$$
The concrete realisation
$$
{\bf H}=2x_5\pa_{x_5},\quad {\bf E}= \sqrt{x_5}\pa_{y_5},\quad {\bf F}=2\sqrt{x_5}\,\pa_{x_5}+\frac{y_5}{\sqrt{x_5}}\, \pa_{y_5},
$$
gives the Laplace-Beltrami operator
$$
L_b= {\cal C} = 2x_5 \pa_{x_5}\pa_{y_5}+y_5\pa_{y_5}^2+\pa_{y_5}.
$$
which is invariant under the {\em involution}
$$
(\bar x_5,\bar y_5)=\left(\frac{y_5^2}{x_5},y_5\right) \quad\mbox{giving the automorphism}\quad
                                     {\bf E}\leftrightarrow {\bf F},\;\; {\bf H}\rightarrow -{\bf H}.
$$
The concrete realisation
$$
{\bf H}=\left(x_8-\frac{3}{4}y_8^2\right)\pa_{x_8}-y_8\pa_{y_8},\quad {\bf E}= y_8\pa_{x_8}+2\pa_{y_8},\quad {\bf F}=\pa_{x_8},
$$
gives the Laplace-Beltrami operator
$$
L_b= {\cal C} = y_8\pa_{x_8}^2+2\pa_{x_8}\pa_{y_8},
$$
which is invariant under the {\em involution}
$$
(\bar x_8,\bar y_8)=\left(\frac{(y_8^2+4x_8)^2}{16}+\frac{y_8}{2},2x_8-\frac{y_8^2}{2}\right) \quad\mbox{giving the automorphism}\quad
                                     {\bf E}\leftrightarrow {\bf F},\;\; {\bf H}\rightarrow -{\bf H}.
$$
These realisations are related through the changes of coordinates: $x_5=\frac{1}{16}\, (4x_8-y_8^2)^2,\; y_5=\frac{y_8}{8}\, (4x_8-y_8^2)$.

\section{Adding Potentials: Separability and Super-Integrability}\label{super}

Second order operators, commuting with $L_b$, are just symmetric quadratic
forms of Killing vectors.  Suppose $\bf K$ is such an operator.  Then we may
seek functions $U$ and $V$, such that
\be  \label{lk=0}   %
[L_b+U,{\bf K}+V]=0,
\ee   %
which constitutes a coupled system of partial differential equations for $U$
and $V$.  The solution depends upon a pair of arbitrary functions, each of one
variable, typical of separable systems.  Requiring that there exist two of
these second order commuting operators, $I_j={\bf K}_j+V_j$, strongly
constrains these (formerly) arbitrary functions, which reduce to rational
functions depending upon only a finite number of parameters.  The coefficients
of the second order derivatives in such an operator $\bf K$ define a
contravariant, rank-two Killing tensor.  For brevity, we will refer to these
operators as Killing tensors in what follows.

Such calculations are standard, so we only present (brief) details of Case I and then just some super-integrable cases for the others.  In the calculations it is only necessary to consider one of each of the equivalence classes listed in Table \ref{symtab}.  Within each equivalence class we can (approximately) halve the number of choices of quadratic Killing tensor, by using the corresponding involutive automorphism.

\subsection{Krall-Sheffer Metrics I and IX}

First consider $I_1={\bf A}^2+V$ in (\ref{lk=0}).  The bracket of two second order operators would generally be {\em third} order, but in this case the third and second order terms automatically vanish, since ${\bf A}^2$ {\em commutes} with $L_b$.  For the remaining 3 coefficients to vanish, we require
$$
(x-1)V_x+yV_y=4(1-x-y)U_x,\quad xV_x+(y-1)V_y=0,\quad L_bV={\bf A}^2U.
$$
The integrability condition for the first pair is just
$$
\pa_x (xU_x+(y-1)U_y+U)=0 ,
$$
leading to the general solution of both $U$ and $V$ depending upon 2 arbitrary functions:
\be  \label{pots1a}  %
U(x,y)=U_1(y)+\frac{U_2\left(\frac{y-1}{x}\right)}{y-1},\quad V(x,y)=4 U_2\left(\frac{y-1}{x}\right).
\ee
The third condition is then automatically satisfied.

\br[Separation of Variables]
Written in terms of the coordinates $\xi=(y-1)/x,\; \eta=y$, the operator $L$ (with the potential (\ref{pots1a})) is separable.
\er  %

The case of $I_2={\bf B}^2+V$ is obtained from $I_1$ by using the involution.  In this particular case, this is a very elementary operation of switching the roles of $x$ and $y$.

The case of $I_3={\bf C}^2+V$ is calculated independently and leads to another second order PDE for $U(x,y)$:
$(x\pa_x+y\pa_y+2) (U_x-U_y)=0$, giving
\be  \label{pots1c}  %
U(x,y)=U_3(x+y)+\frac{U_4\left(\frac{y}{x}\right)}{x+y},\quad V(x,y)=-4 U_4\left(\frac{y}{x}\right),
\ee
which is symmetric (in form) under the involution.

\subsubsection{Super-Integrability}

Requiring two of above operators (say $I_1$ and $I_2$) to commute with $L$ reduces the potential to a rational function.
In this case we have {\em three} such commuting operators, $L=L_b+U,\; I_1=A^2+V_1,\; I_2=B^2+V_2,\; I_3=C^2+V_3$, where
\be\label{superpots1}  %
\begin{array}{ll}
\displaystyle U=\frac{k_1}{(1-x-y)}+\frac{k_2}{x}+\frac{k_3}{y},
    & \displaystyle   V_1=\frac{4k_1x}{(x+y-1)}+\frac{4k_2(y-1)}{x},\\[3mm]
\displaystyle V_2=\frac{4k_1y}{(x+y-1)}+\frac{4k_3(x-1)}{y},
    & \displaystyle    V_3=-\, \frac{4k_2y}{x}-\, \frac{4k_3x}{y}.
\end{array}
\ee  %
These operators are not independent, satisfying $4L+I_1+I_2+I_3=4k_1$, which reduces to $4L_b+A^2+B^2+C^2=0$, when the parameters are zero.  If we extend the involution to act on the parameters as $k_1\rightarrow k_1,\; k_2\leftrightarrow k_3$, then
$$
L\rightarrow L,\quad I_1 \leftrightarrow I_2,\quad I_3 \rightarrow I_3.
$$

\subsubsection{The Transformation to Case IX}

We have a transformation between the Laplace-Beltrami operators of cases I and IX.  The potential (\ref{pots1a}) is transformed to
$$
U(x,y)=U_1(y^2)+\frac{U_2\left(\frac{y^2-1}{x^2}\right)}{y^2-1},\quad V(x,y)=4 U_2\left(\frac{y^2-1}{x^2}\right).
$$
The potential (\ref{pots1c}) is transformed to
$$
U(x,y)=U_3(x^2+y^2)+\frac{U_4\left(\frac{y^2}{x^2}\right)}{x+y},\quad V(x,y)=-4 U_4\left(\frac{y^2}{x^2}\right).
$$
In particular, the potentials (\ref{superpots1}) give us the corresponding super-integrable case for Case IX.

\subsubsection{The Krall-Sheffer Gauge}

Starting with the operator $L=L_b+U$, with $U$ defined by (\ref{superpots1}), the gauge transformation
$$
L\mapsto G^{-1}LG,\quad\mbox{with}\quad G=(1-x-y)^ax^by^c
$$
leads to $k_1=\frac{1}{2}a(2a-1),\, k_2=\frac{1}{2}b(2b-1),\, k_3=\frac{1}{2}c(2c-1)$ if we require the potential to reduce to a constant.  The operator then reduces to the Krall-Sheffer form with
$$
a=\frac{1}{4}(2(\beta+\kappa_1+\kappa_2)-1),\quad b=-\frac{1}{4}(2\kappa_1+1),\quad c=-\frac{1}{4}(2\kappa_2+1).
$$
The operators $I_1, I_2$ and $I_3$ then (up to an additive constant) take the form
\bea  %
 I_1 &=& x(1-x-y)\pa_x^2+(\kappa_1(y-1)-(\beta+\kappa_2)x)\pa_x,\nn\\
  I_2 &=& y(1-x-y)\pa_y^2+(\kappa_2(x-1)-(\beta+\kappa_1)y)\pa_y, \nn\\
  I_3 &=& xy(\pa_x-\pa_y)^2+(\kappa_2x-\kappa_1y)(\pa_x-\pa_y).  \nn
\eea  %

\br
The choice of variables entering the definition of $G$ is dictated by the potential $U$, but the specific form needs to be determined.
\er

If we transform to case IX the second order parts of the two $L$ operators are related in the usual way, but the first order parts take the form
$$
\frac{2 \kappa_1+1+(2\beta-1)x^2}{x}\, \pa_x+ \frac{2 \kappa_2+1+(2\beta-1)y^2}{y}\, \pa_y,
$$
containing non-polynomial parts.  To remove these we must set $\kappa_1=\kappa_2=-\frac{1}{2}$.  The other coefficients are related by $\beta_9=2\beta_1-1$.  It should be noted that this gives the {\em full} parameter version of Case IX from a restricted version of Case I.

\subsection{Super-Integrability for Metrics II and III}

If we consider $L=L_b+U,\; I_1=H^2+V_1,\; I_2=E^2+V_2$, then we find
\be\label{superpots2he}  %
U=\frac{k_1}{x}+\frac{k_2}{y}+\frac{k_3(y-1)}{x^2},\quad
 V_1=\frac{16k_1(1-y)}{x}-\frac{16k_3(1-y)^2}{x^2},\quad V_2=-4k_2\, \frac{x}{y}-4k_3\, \frac{y}{x}.
\ee  %

\br
The commutator $[I_1,I_2]$ is a third order operator, so cannot be written as a polynomial expression in $L,\, I_1,\, I_2$.  However, the next commutators, $[I_1,[I_1,I_2]]$ and $[I_2,[I_1,I_2]]$ can, giving us a quadratic algebra (see \cite{f07-1}), but this will not be considered here.
Such operator algebras play an important role in the theory of exactly and quasi-exactly solvable operators \cite{09-5}.
\er

Under the involution, these operators transform to $\bar L=L_b+\bar U,\; \bar I_1=H^2+\bar V_1,\; I_3=F^2+V_3$, where
$$
\bar U=\frac{k_1x}{(y-1)^2}+\frac{k_2}{y}+\frac{k_3x^2}{(y-1)^3},\quad
\bar V_1=\frac{16k_1x}{(1-y)}-\frac{16k_3x^2}{(1-y)^2},\quad V_3=-4k_2\, \frac{(y-1)^2}{xy}-4k_3\, \frac{xy}{(y-1)^2}.
$$

If we consider $L=L_b+U,\; I_2=E^2+V_2,\; I_3=F^2+V_3$, then we find
$$
U=\frac{k_2}{y},\quad V_2= \frac{-4k_2x}{y}, \quad V_3=\frac{-4k_2(y-1)^2}{xy},
$$
and, furthermore, $[H,L]=0$.  These operators all commute with $L$ and form the algebra:
$$
 [H,I_2]=4I_2,\quad  [H,I_3]=-4I_3,\quad [I_2,I_3]=16 HL-H^3+2(8k_2-1) H.
$$
They also satisfy the constraint:
$$
I_2I_3+I_3I_2=32L^2-16(4k_2+1)L-4H^2L+\frac{1}{8}H^4+\frac{1}{2} (5-8k_2)H^2+16k_2(2k_2-1).
$$

\subsubsection{The Transformation to Case III}

All of these formula can be transformed to Case III.  For example, this last algebra is satisfied by
$$
L=L_b-\frac{4k_2x}{y^2},\quad I_2=256\left(E^2-\frac{4k_2x^2}{y^2}\right),\quad I_3=\frac{1}{256}\left(F^2-\frac{4k_2x^2}{y^2}\right), \quad H,
$$
where $L_b,\, E,\, F$ and $H$ are those in Case III.
The numerical factor follows from the fact that $E_2\mapsto -16 E_3,\;\; F_2\mapsto -\frac{1}{16} F_3,\;\; H_2\mapsto H_3$, which satisfy the same commutation relations, since such a scaling is just another isomorphism of the algebra.

\subsubsection{The Krall-Sheffer Gauge}

Starting with the operator $L=L_b+U$, with $U$ defined by (\ref{superpots2he}), the gauge transformation
$$
L\mapsto G^{-1}LG,\quad\mbox{with}\quad G=e^{a(y-1)/x}x^by^c
$$
leads to $k_1=a(1-2b),\, k_2=\frac{1}{2}c(2c-1),\, k_3=a^2$ if we require the potential to reduce to a constant.  The operator then reduces to the Krall-Sheffer form with
$$
a=\frac{\kappa_1}{2},\quad b=\frac{1}{2}(\beta+\kappa_2-1),\quad c=-\frac{1}{4}(2\kappa_2+1).
$$
The operators $I_1$ and $I_2$ then (up to an additive (and an overall multiplicative) constant) take the form
$$
 I_1 = x^2\pa_x^2+((\beta+\kappa_2)x+\kappa_1(1-y))\pa_x,   \quad
  I_2 = xy\pa_y^2+(\kappa_1y-\kappa_2x)\pa_y.
$$

\paragraph{Transforming to case III:} the second order parts of the two $L$ operators are related in the usual way, but the first order parts take the form
$$
\left(\beta x-\frac{\kappa_1}{64}\right)\, \pa_x+ \left(\beta y-\frac{\kappa_1y}{128 x}-(2\kappa_2+1)\frac{x}{y}\right)\, \pa_y,
$$
containing non-polynomial parts.  To remove these we must set $\kappa_1=0, \kappa_2=-\frac{1}{2}$, which leaves us with the reduced operator with first order part given by $\beta x\pa_x+\beta y\pa_y$.

To obtain the most general Krall-Sheffer operator we need to consider the operators $\frac{1}{4}(HE+EH)$ and $E^2$.  We find the following result:
$$
L=L_b+\frac{k_1}{x}+\frac{k_2y}{x^2}+\frac{k_3(x+y^2)}{x^3}
$$
commutes with the two operators
$$
I_1=\frac{1}{4}(HE+EH)-\frac{k_1y}{x}-\frac{k_2(x+y^2)}{x^2}-\frac{k_3y(2x+y^2)}{x^3}\quad\mbox{and}\quad I_2=E^2+\frac{k_2y}{x}+\frac{k_3y^2}{x^2}.
$$
The gauge transformation
$$
L\mapsto G^{-1}LG,\quad\mbox{with}\quad G=x^a e^{b(2x+y^2)/x^2+cy/x}
$$
leads to $k_1=b(4a-3)-c^2,\, k_2=-4bc,\, k_3=-4b^2$ if we require the potential to reduce to a constant.  The operator then reduces to the Krall-Sheffer form with
$$
a=\frac{\kappa_1}{2},\quad b=\frac{1}{2}(\beta+\kappa_2-1),\quad c=-\frac{1}{4}(2\kappa_2+1).
$$
The operators $I_1$ and $I_2$ then take the form
$$
 I_1 = 2x^2\pa_x\pa_y+xy\pa_y^2+(\kappa_2x-\kappa_1y)\pa_x+(\beta x+\kappa_1)\pa_y,   \quad
  I_2 = x^2\pa_y^2+(\kappa_2x-\kappa_1y)\pa_y.
$$

\subsection{The Metrics IV, VI and VII}

These cases are separated in Cartesian coordinates, so the examples of super-integrable cases are very simple and well known.  In the Krall-Sheffer gauge the operators $L$ are just direct sums of $1-$dimensional ones, corresponding to products of classical polynomials, so we omit these examples here.

\subsection{Super-Integrability for Metrics V and VIII}

If we consider $L=L_b+U,\; I_1=H^2+V_1,\; I_2=E^2+V_2$, then we find
\be\label{superpots5he}  %
U=k_1 y+\frac{k_2}{x}+\frac{k_3 y}{x^2},\quad
 V_1=-\,\frac{k_2 y}{x}-\,\frac{k_3 y^2}{x^2},\quad V_2= k_1 x-\, \frac{k_3}{x}.
\ee  %
The involution transforms to the case with $L=L_b+\bar{U},\; I_1=H^2+\bar{V}_1,\; I_3=F^2+\bar{V}_3$.

We may also consider the case $L=L_b+U,\; I_2=E^2+V_2,\; I_3=F^2+V_3$, which gives
$$
U=k_1 y+ k_2\sqrt{x}+\frac{k_3 y}{\sqrt{x}},\quad
 V_2= k_1 x +2 k_3 \sqrt{x},\quad V_3= \frac{k_1 y^2}{x}+ \frac{2k_2 y}{\sqrt{x}},
$$
which is invariant under the involution.

\subsubsection{The Krall-Sheffer Gauge}

Starting with the case (\ref{superpots5he}), the gauge transformation
$$
L\mapsto G^{-1}LG,\quad\mbox{with}\quad G=e^{(a x+b)y/x}x^c
$$
leads to $k_1=-a^2,\, k_2=b(1-2c),\, k_3=b^2$ if we require the potential to reduce to a constant.  The operator then reduces to the Krall-Sheffer form with
$$
a=\frac{\beta}{2},\quad b=\frac{\kappa_1}{2},\quad c=-\frac{(\kappa_2-1)}{2}.
$$
The operators $I_1$ and $I_2$ then (up to an additive (and an overall multiplicative) constant) take the form
$$
 I_1 = x^2\pa_x^2+(\kappa_2x-\kappa_1y)\pa_x,   \quad
  I_2 = x\pa_y^2+(\beta x+\kappa_1)\pa_y.
$$

\paragraph{Case VIII:} Transforming between these cases restricts the parameters to the Laplace-Beltrami case.
Here we need the choice $L=L_b+U,\; I_1=F^2+V_1,\; I_2=\frac{1}{4}(E^2-2(HF+FH))+V_2$, where
\bea  %
&& U= \frac{\beta^2}{4} y(y^2-2x)+\frac{\beta \kappa_2}{2}(y^2-x)+\frac{y}{4}(\kappa_2^2-2\beta \kappa_1),\nn\\
&& V_1= -\frac{1}{4}(\beta y+\kappa_2)^2,\quad V_2= \frac{\beta^2}{4} (y^2-x)x + \frac{x}{4}(\kappa_2^2+2\beta (\kappa_2 y-\kappa_1)).\nn
\eea  %
Using $G= e^{(6\kappa_2 x+6\kappa_1 y+6\beta xy-3\kappa_2 y^2-2 \beta y^3)/12}$, these gauge transform (up to additive constants) to:
\bea  %
L &=& y\pa_x^2+2\pa_x\pa_y+(\beta x+\kappa_1)\pa_x+(\beta y+\kappa_2)\pa_y,\nn\\
I_1 &=& \pa_x^2+(\beta y+\kappa_2)\pa_x,  \label{8ks}\\
I_2 &=& (y^2-x)\pa_x^2+2y\pa_x\pa_y+\pa_y^2+(\kappa_1 y-\kappa_2 x)\pa_x+(\beta x+\kappa_1)\pa_y,\nn
\eea  %

\section{The Polynomial Eigenfunctions}\label{pols}

By Definition \ref{def:admiss}, an {\em admissible operator} possesses a
sequence of eigenvalues $\lambda_N$ (with $\lambda_N\neq\lambda_M$ for $M\neq N$), such that, for each $N\ge 0$, there
exist $N+1$ linearly independent polynomial eigenfunctions of degree $N$, with
eigenvalue $\lambda_N$.  The {\em degeneracy} (for $N>0$) of $\lambda_N$ stems
from the super-integrability in a very explicit way.  We show below how to use
the operators $I_1$ and $I_2$ to build sequences of polynomial eigenfunctions
of $L$.

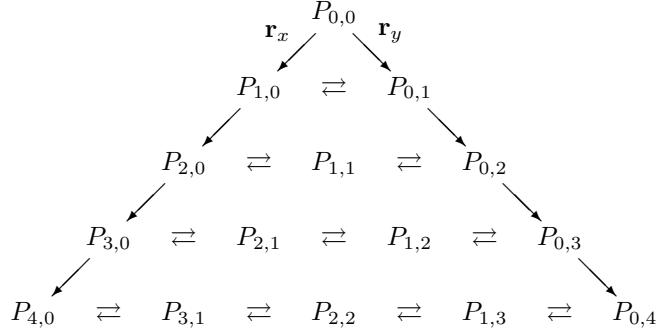
\begin{figure}[hbt]
\begin{center}
\unitlength=0.5mm
\begin{picture}(160,80)
\put(80,80){\makebox(0,0){$P_{0,0}$}}
\put(60,60){\makebox(0,0){$P_{1,0}$}}
\put(100,60){\makebox(0,0){$P_{0,1}$}}
\put(40,40){\makebox(0,0){$P_{2,0}$}}
\put(80,40){\makebox(0,0){{$P_{1,1}$}}}
\put(120,40){\makebox(0,0){$P_{0,2}$}}
\put(20,20){\makebox(0,0){$P_{3,0}$}}
\put(60,20){\makebox(0,0){$P_{2,1}$}}
\put(100,20){\makebox(0,0){$P_{1,2}$}}
\put(140,20){\makebox(0,0){$P_{0,3}$}}
\put(0,0){\makebox(0,0){$P_{4,0}$}}
\put(40,0){\makebox(0,0){{$P_{3,1}$}}}
\put(80,0){\makebox(0,0){$P_{2,2}$}}
\put(120,0){\makebox(0,0){{$P_{1,3}$}}}
\put(160,0){\makebox(0,0){$P_{0,4}$}}
\put(75,75){\vector(-1,-1){10}}
\put(85,75){\vector(1,-1){10}}
\put(55,55){\vector(-1,-1){10}}  \put(105,55){\vector(1,-1){10}}
\put(35,35){\vector(-1,-1){10}}  \put(125,35){\vector(1,-1){10}}
\put(15,15){\vector(-1,-1){10}}  \put(145,15){\vector(1,-1){10}}
\put(80,60){\makebox(0,0){$\rightleftarrows$}}
\put(60,40){\makebox(0,0){$\rightleftarrows$}}
\put(100,40){\makebox(0,0){$\rightleftarrows$}}
\put(40,20){\makebox(0,0){$\rightleftarrows$}}
\put(80,20){\makebox(0,0){$\rightleftarrows$}}
\put(120,20){\makebox(0,0){$\rightleftarrows$}}
\put(20,0){\makebox(0,0){$\rightleftarrows$}}
\put(60,0){\makebox(0,0){$\rightleftarrows$}}
\put(100,0){\makebox(0,0){$\rightleftarrows$}}
\put(140,0){\makebox(0,0){$\rightleftarrows$}}
\put(65,75){\makebox(0,0){${\bf r}_x$}}
\put(95,75){\makebox(0,0){${\bf r}_y$}}
\end{picture}
\end{center}
\caption{The triangular lattice of polynomials $P_{m,n}$
with $P_{0,0}=1$.  Horizontal arrows denote the
action of $I_1$ (right) and $I_2$ (left). The operators ${\bf r}_x$ and ${\bf r}_y$ represent the edge ladder operators.} \label{trilatt}
\end{figure}

We may choose a basis of {\em monic} polynomials $P_{m,n}$ of the form (\ref{pmn})
and organise them in a triangular array (see Figure \ref{trilatt}), with the degree $N=m+n$ constant on horizontal layers.
Since the eigenvalue (\ref{lambda}) depends only on the degree, it is constant on each horizontal level.

\br  %
Cases $IV, VI$ and $VII$ separate into sums of elementary $1-$dimensional operators so their eigenfunctions are just products of classical orthogonal polynomials (respectively Laguerre-Laguerre, Laguerre-Hermite and Hermite-Hermite), so will not be discussed further.
\er  %

The Krall-Sheffer operators reduce to one dimensional operators on at least one of the edges of this triangle (usually both).  The two exceptional cases are III, which only reduces on the left and VIII, which only reduces on the right.  This means that there exist {\em 3 point recurrence relations} for the eigenfunctions $P_{m,0}$ and/or $P_{0,n}$.  Below, we give explicit {\em 3 {\bf level} recurrence relations} for each case and show how these reduce to the {\em 3 point recurrence relations} on the appropriate edges.  The one dimensional reductions on the left and/or right edges also possess {\em differential ladder operators}.  We also present the ladder operators in directions which are ``parallel'' to the edges for each case and show how these reduce to their one dimensional counterparts on appropriate edges.

\paragraph{What is the role of the commuting operators?} Since
$$
\left. \begin{array}{l}
Lf=\lambda f \\[2mm]
[I,L]=0
\end{array}\right\}\quad\Rightarrow\quad L(If)=\lambda (If),
$$
the function $If$ is also an eigenfunction with the same eigenvalue.  For our Krall-Sheffer operators, we have two such operators $I_1, I_2$ which take us respectively right and left on each horizontal level.  Starting from $P_{N,0}$ or $P_{0,N}$, we use these operators to build the eigenfunctions $P_{m,n}$, with $N=m+n$.

\paragraph{The $3$ Level Recurrence and Ladder Operators.} It is possible to explicitly determine the coefficients of the 3 level recurrence relations (\ref{ks3level}).  This is a laborious calculation, so only the resulting recurrence relations are presented.  The general form of (\ref{ks3level}) constitutes a difference equation on $2(m+n+1)$ points, but our formulae contain at most $9$ points, and each equation fits into the following $12$ point relation:
\bea  %
P_{m+1,n} &=& x P_{m,n} +\sum_{i=0}^3 a_i P_{m-1+i,n+1-i}+\sum_{i=0}^{6}b_i P_{m-3+i,n+2-i}, \nn\\[-1.5mm]
&&  \label{12point}  \\
P_{m,n+1} &=& y P_{m,n} +\sum_{i=0}^{m+n} c_i P_{m+1-i,n-1+i}+\sum_{i=0}^6 d_i P_{m+2-i,n-3+i}, \nn
\eea  %
{\em Given} the first of these, we operate with
$$
L=A\pa_x^2+2B\pa_x\pa_y+C\pa_y^2+D\pa_x+E\pa_y,
$$
taking into account that $P_{m,n}$, is an eigenfunction, with eigenvalue $\lambda_{m+n}$, we find
\bea  %
\lambda_{m+n+1}P_{m+1,n} &=& \lambda_{m+n} (x P_{m,n}+\sum_{i=0}^3a_i P_{m-1+i,n+1-i})+(2A\pa_x+2B\pa_y+D)P_{m,n}\nn\\
&& \qquad\qquad +\lambda_{m+n-1} \sum_{i=0}^{6}b_i P_{m-3+i,n+2-i}.  \nn
\eea  %
Subtracting $\lambda_{m+n-1}$ times the first of (\ref{12point}), leads to a first order {\em two level} differential-difference relation:
\bea  %
(\lambda_{m+n+1}-\lambda_{m+n-1})P_{m+1,n} &=& (\lambda_{m+n}-\lambda_{m+n-1}) (x P_{m,n}+\sum_{i=0}^3a_i P_{m-1+i,n+1-i}) \nn\\
&& \qquad\qquad +(2A\pa_x+2B\pa_y+D)P_{m,n}.  \label{2leveldiffdel}
\eea  %
The next step in the calculation is to use the {\em horizontal shift} operators $I_1$ and $I_2$ to write all the polynomials of degree $m+n$ (arising in (\ref{2leveldiffdel})) in terms of the polynomial $P_{m,n}$.  In principle, this could result in a fourth order differential operator, but only second and first order operators occur.  This, together with a similar manipulation of the second of (\ref{12point}), gives two differential operators which {\em raise the degree} of the polynomial eigenfunction:
\be\label{r+}  %
R_{+,x}P_{m,n}=P_{m+1,n},\quad R_{+,y}P_{m,n}=P_{m,n+1}.
\ee  %
It is also possible to construct {\em lowering operators}, but these are usually considerably more complicated and
(usually) depend explicitly upon both $m$ and $n$ (not just through the combination $m+n$).

The formulae for $[L,R_{+,x}]$ and $[L,R_{+,y}]$ are non-standard but still imply that $R_{+,x}P_{m,n}$ and $R_{+,y}P_{m,n}$ are eigenfunctions of $L$, with eigenvalue $\lambda_{m+n+1}$.

\subsection{Krall-Sheffer Metric I}\label{pols1}

Here we have
\bea  %
L &=& (x^2-x) \pa_x^2+2 x y \pa_x\pa_y+ (y^2-y) \pa_y^2+ (\beta x+\kappa_1)\pa_x+ (\beta y+\kappa_2)\pa_y, \nn\\
 I_1 &=& x(1-x-y)\pa_x^2+(\kappa_1(y-1)-(\beta+\kappa_2)x)\pa_x,\nn\\
  I_2 &=& y(1-x-y)\pa_y^2+(\kappa_2(x-1)-(\beta+\kappa_1)y)\pa_y, \nn\\
  I_3 &=& xy(\pa_x-\pa_y)^2+(\kappa_2x-\kappa_1y)(\pa_x-\pa_y).  \nn
\eea  %
The polynomial eigenfunctions of $L$ satisfy the $6$ point difference equations:
\bea  %
P_{m+1,n} &=& x P_{m,n}  \nn\\
&& +\frac{1}{\gamma_{2N}\gamma_{2N-2}}\, (((\beta+2n-2)(\kappa_1-2m)-2m(m+1)) P_{m,n} + 2n(n-\kappa_2-1) P_{m+1,n-1}) \nn\\
&& +\frac{1}{\gamma_{2N-1}\gamma_{2N-2}^2\gamma_{2N-3}}\, (m(\beta+m+2n-2)(\kappa_1-m+1)(\beta+\kappa_1+m+2n-1) P_{m-1,n} \nn\\ && +n(\kappa_2-n+1)((\beta+2n-3)(\kappa_1-2m)-2m(m+1)) P_{m,n-1}  \nn\\
&& \hspace{2cm} -n(n-1)(\kappa_2-n+1)(\kappa_2-n+2)P_{m+1,n-2}),  \label{1pmp1n}
\eea  %
and
\bea  %
P_{m,n+1} &=& y P_{m,n}  \nn\\
&&   +\frac{1}{\gamma_{2N}\gamma_{2N-2}}\,
   (((\beta+2m-2)(\kappa_2-2n)-2n(n+1))P_{m,n} + 2m(m-\kappa_1-1) P_{m-1,n+1})  \nn\\
&& +\frac{1}{\gamma_{2N-1}\gamma_{2N-2}^2\gamma_{2N-3}}\, (n(\beta+2m+n-2)(\kappa_2-n+1)(\beta+\kappa_2+2m+n-1)P_{m,n-1}  \nn\\
&& +m(\kappa_1-m+1)((\beta+2m-3)(\kappa_2-2n)-2n(n+1))P_{m-1,n}  \nn\\
&&  \hspace{2cm} -m(m-1)(\kappa_1-m+1)(\kappa_1-m+2) P_{m-2,n+1}),  \label{1pmnp1}
\eea  %
where $N=m+n, \gamma_n=\beta+n$.  The $6$ points of these difference equations are depicted by bullet points in Figure \ref{diff1fig}.
\begin{figure}[htb]
\begin{center}
\unitlength=0.3mm
\begin{picture}(160,60)
\put(60,60){\makebox(0,0){$\circ$}}
\put(50,50){\makebox(0,0){$\circ$}}
\put(70,50){\makebox(0,0){$\circ$}}
\put(40,40){\makebox(0,0){$\circ$}}
\put(60,40){\makebox(0,0){{$\circ$}}}
\put(80,40){\makebox(0,0){$\circ$}}
\put(30,30){\makebox(0,0){$\circ$}}
\put(50,30){\makebox(0,0){$\circ$}}
\put(70,30){\makebox(0,0){$\circ$}}
\put(90,30){\makebox(0,0){$\circ$}}
\put(20,20){\makebox(0,0){$\circ$}}
\put(40,20){\makebox(0,0){$\bullet$}}
\put(60,20){\makebox(0,0){$\bullet$}}
\put(80,20){\makebox(0,0){$\bullet$}}
\put(100,20){\makebox(0,0){$\circ$}}
\put(10,10){\makebox(0,0){{$\circ$}}}
\put(30,10){\makebox(0,0){{$\circ$}}}
\put(50,10){\makebox(0,0){{$\bullet$}}}
\put(70,10){\makebox(0,0){{$\bullet$}}}
\put(90,10){\makebox(0,0){{$\circ$}}}
\put(110,10){\makebox(0,0){{$\circ$}}}
\put(0,0){\makebox(0,0){$\circ$}}
\put(20,0){\makebox(0,0){{$\circ$}}}
\put(40,0){\makebox(0,0){$\circ$}}
\put(60,0){\makebox(0,0){$\bullet$}}
\put(80,0){\makebox(0,0){$\circ$}}
\put(100,0){\makebox(0,0){$\circ$}}
\put(120,0){\makebox(0,0){$\circ$}}
\end{picture}
\end{center}
\caption{The pattern for the difference equations for both $P_{m+1,n}$ and $P_{m,n+1}$ for both cases I and II.}\label{diff1fig}
\end{figure}
Notice that when $n=1$, the point $(m+1,n-2)$ is out of the Figure, but also that in formula (\ref{1pmp1n}) the coefficient of $P_{m+1,n-2}$ vanishes when $n=1$, so the formula is still valid.  When $n=0$, the three points $(m+1,n-2), (m+1,n-1)$ and $(m,n-1)$ fall outside the Figure, but again, the corresponding coefficients vanish, leaving us with the {\em 3 {\bf point} recurrence relation} on the left edge:
\bea  %
P_{m+1,0} &=& \left(x+\frac{(\beta-2)(\kappa_1-2m)-2m(m+1)}{(\beta+2m)(\beta+2m-2)}\right) P_{m,0}  \nn\\
&& +\left(\frac{m(\beta+m-2)(\kappa_1-m+1)(\beta+\kappa_1+m-1)}{(\beta+2m-1)(\beta+2m-2)^2(\beta+2m-3)}\right) P_{m-1,0}.  \label{1pmp10}
\eea  %
Similarly, as we approach the right edge, the formula (\ref{1pmnp1}) consistently reduces to the {\em 3 point recurrence relation} on the right edge:
\bea  %
P_{0,n+1} &=& \left(y+\frac{(\beta-2)(\kappa_2-2n)-2n(n+1)}{(\beta+2n)(\beta+2n-2)}\right) P_{0,n}  \nn\\
&& +\left(\frac{n(\beta+n-2)(\kappa_2-n+1)(\beta+\kappa_2+n-1)}{(\beta+2n-1)(\beta+2n-2)^2(\beta+2n-3)}\right) P_{0,n-1}.  \label{1p0np1}
\eea  %
The left and right edge polynomials satisfy respective Jacobi equations:
\bea  %
L^{(x)} P_{m,0}=(x(x-1)\pa_x^2+(\beta x+\kappa_1)\pa_x)P_{m,0} &=& m(m-1+\beta)P_{m,0}, \nn\\
L^{(y)} P_{0,n}=(y(y-1)\pa_y^2+(\beta y+\kappa_2)\pa_y)P_{0,n} &=& n(n-1+\beta)P_{0,n}.  \nn
\eea  %
To proceed with these {\em second order} difference equations, we expect to need the three lowest polynomials:
\be\label{p00}  %
P_{0,0}=1,\quad P_{1,0}= x+\frac{\kappa_1}{\beta}, \quad P_{0,1}= y+\frac{\kappa_2}{\beta},
\ee  %
which are calculated directly as monic, degree one polynomial eigenfunctions. However, when $m=0$, (\ref{1pmp10}) is consistent and can be used to construct $P_{1,0}$ (and similarly for $P_{0,1}$).  We then use the difference equations (\ref{1pmp1n}) and (\ref{1pmnp1}) to calculate the degree 2, degree 3, etc polynomials successively.

Alternatively, we may calculate the polynomials $P_{m,0}$ and $P_{0,n}$, using (\ref{1pmp10}) and (\ref{1p0np1}) and then use the commuting operators to calculate the remaining polynomials on each level.  We know that, since the operators $I_1$ and $I_2$ commute with $L$, the functions $I_kP_{m,n}, k=1,2$, are also polynomial eigenfunctions of the same degree.  By looking at the leading order (monic) part, we obtain:
\bea  %
I_1P_{m,n}+m(\beta+\kappa_2+m-1)P_{m,n} &=& m(\kappa_1-m+1)P_{m-1,n+1} ,\nn\\
I_2P_{m,n}+n(\beta+\kappa_1+n-1)P_{m,n} &=& n(\kappa_2-n+1)P_{m+1,n-1} ,\nn
\eea  %
We see that, starting with $P_{m,0}$, $I_1$ moves us to the right, until we reach $P_{0,n}$, which satisfies $I_1P_{0,n}=0$.  $I_2$ similarly moves us from right to left.

Finally, we may construct the operators (\ref{r+}), which are given by
\bea  %
R_{+,x} &=& \frac{1}{(\beta+2N)(\beta+2N-1)}((\beta+N-1)((\beta+2N)x+\kappa_1-N)+(\beta+2N)x(x-1)\pa_x  \nn\\
  &&  +((\beta+2N)xy+(\beta+\kappa_1)y+\kappa_2(1-x))\pa_y+y(x+y-1)\pa_y^2),  \nn\\
R_{+,y} &=& \frac{1}{(\beta+2N)(\beta+2N-1)}((\beta+N-1)((\beta+2N)y+\kappa_2-N)+(\beta+2N)y(y-1)\pa_y  \nn\\
  &&  +((\beta+2N)xy+(\beta+\kappa_2)x+\kappa_1(1-y))\pa_x+x(x+y-1)\pa_x^2).  \nn
\eea  %
The formulae interchange under the transformation $x\leftrightarrow y, \kappa_1\leftrightarrow \kappa_2$.
These operators satisfy the commutation relations
\bea  %
[L,R_{+,x}] &=& \frac{(2x-1)(L-\lambda_N)}{\beta+2N-1} +(\lambda_{N+1}-\lambda_N)R_{+,x}, \nn\\[2mm]
[L,R_{+,y}] &=& \frac{(2y-1)(L-\lambda_N)}{\beta+2N-1} +(\lambda_{N+1}-\lambda_N)R_{+,y}. \nn
\eea  %

Starting from $P_{0,n}$, $R_{+,x}$ is used to build the functions $P_{m,n}, m\geq 1$, with $n$ fixed.  When $n=0$, $R_{+,x}$ reduces to the $1$ dimensional ladder operator
$$
r_{+,x} = \frac{1}{(\beta+2m)(\beta+2m-1)}((\beta+m-1)((\beta+2m)x+\kappa_1-m)+(\beta+2m)x(x-1)\pa_x).
$$
Similar statements can be made regarding $R_{+,y}$.

This particular case is symmetric in $(x,y)$ and $(\kappa_1,\kappa_2)$, which is reflected in all the formulae.

\subsection{Krall-Sheffer Metric II}\label{pols2}

Here we have
\bea  %
L &=& x^2 \pa_x^2+2 x y \pa_x\pa_y+ (y^2-y) \pa_y^2+ (\beta x+\kappa_1)\pa_x+ (\beta y+\kappa_2)\pa_y, \nn\\
 I_1 &=& x^2\pa_x^2+((\beta+\kappa_2)x+\kappa_1(1-y))\pa_x,\nn\\
  I_2 &=& xy\pa_y^2+(\kappa_1y-\kappa_2x)\pa_y. \nn
\eea  %
The polynomial eigenfunctions of $L$ satisfy the $6$ point difference equations:
\bea  %
P_{m+1,n} &=& x P_{m,n}  \nn\\
&& +\frac{1}{\gamma_{2N}\gamma_{2N-2}}\, (\kappa_1(\beta+2n-2) P_{m,n} + 2n(n-\kappa_2-1) P_{m+1,n-1}) \nn\\
&& +\frac{1}{\gamma_{2N-1}\gamma_{2N-2}^2\gamma_{2N-3}}\, (m\kappa_1^2(\beta+m+2n-2) P_{m-1,n} \nn\\
&& +n\kappa_1(\beta+2n-3)(\kappa_2-n+1) P_{m,n-1}  \nn\\
&& \hspace{2cm} -n(n-1)(\kappa_2-n+1)(\kappa_2-n+2)P_{m+1,n-2}),  \label{2pmp1n}
\eea  %
and
\bea  %
P_{m,n+1} &=& y P_{m,n}  \nn\\
&&   +\frac{1}{\gamma_{2N}\gamma_{2N-2}}\,
   (((\beta+2m-2)(\kappa_2-2n)-2n(n+1))P_{m,n} -2m\kappa_1 P_{m-1,n+1})  \nn\\
&& +\frac{1}{\gamma_{2N-1}\gamma_{2N-2}^2\gamma_{2N-3}}\, (n(\beta+2m+n-2)(\kappa_2-n+1)(\beta+\kappa_2+2m+n-1))P_{m,n-1}  \nn\\
&& +m\kappa_1((\beta+2m-3)(\kappa_2-2n)-2n(n+1))P_{m-1,n} -m(m-1)\kappa_1^2 P_{m-2,n+1}),  \label{2pmnp1}
\eea  %
where $N=m+n, \gamma_n=\beta+n$. The $6$ points of these difference equations are again depicted by the bullet points in Figure \ref{diff1fig}.
Again, as we approach the left edge, the formula (\ref{2pmp1n}) consistently reduces to the {\em 3 point recurrence relation}:
\bea  %
P_{m+1,0} &=& \left(x+\frac{\kappa_1(\beta-2)}{(\beta+2m)(\beta+2m-2)}\right) P_{m,0} \nn\\
&& +\left(\frac{m\kappa_1^2(\beta+m-2)}{(\beta+2m-1)(\beta+2m-2)^2(\beta+2m-3)}\right) P_{m-1,0}.  \label{2pmp10}
\eea  %
Similarly, as we approach the right edge, the formula (\ref{2pmnp1}) consistently reduces to the {\em 3 point recurrence relation}:
\bea  %
P_{0,n+1} &=& \left(y+\frac{(\beta-2)(\kappa_2-2n)-2n(n+1)}{(\beta+2n)(\beta+2n-2)}\right) P_{0,n}  \nn\\
&& +\left(\frac{n(\beta+n-2)(\kappa_2-n+1)(\beta+\kappa_2+n-1)}{(\beta+2n-1)(\beta+2n-2)^2(\beta+2n-3)}\right) P_{0,n-1}.  \label{2p0np1}
\eea  %
The left and right edge polynomials satisfy respectively:
\bea  %
L^{(x)} P_{m,0}=(x^2\pa_x^2+(\beta x+\kappa_1)\pa_x)P_{m,0} &=& m(m-1+\beta)P_{m,0}, \nn\\
  &&  \label{LxLy2}  \\
L^{(y)} P_{0,n}=(y(y-1)\pa_y^2+(\beta y+\kappa_2)\pa_y)P_{0,n} &=& n(n-1+\beta)P_{0,n}.  \nn
\eea  %
Again, we have the polynomials (\ref{p00}), which are the seeds for the difference equations (\ref{2pmp1n}) and (\ref{2pmnp1}) when calculating the degree 2, degree 3, etc polynomials.

Alternatively, we may calculate the polynomials $P_{m,0}$ and $P_{0,n}$, using (\ref{2pmp10}) and (\ref{2p0np1}) and then use the commuting operators to calculate the remaining polynomials on each level.  By looking at the leading order (monic) part of $I_kP_{m,n}$, we obtain:
\bea  %
I_1P_{m,n}-m(\beta+\kappa_2+m-1)P_{m,n} &=& -m\kappa_1P_{m-1,n+1} ,\nn\\
I_2P_{m,n}-n\kappa_1P_{m,n} &=& -n(\kappa_2-n+1)P_{m+1,n-1} ,\nn
\eea  %
We see that, starting with $P_{m,0}$, $I_1$ moves us to the right, until we reach $P_{0,n}$, which satisfies $I_1P_{0,n}=0$.  $I_2$ similarly moves us from right to left.

Finally, we may construct the operators (\ref{r+}), which are given by
\bea  %
R_{+,x} &=& \frac{1}{(\beta+2N)(\beta+2N-1)}((\beta+N-1)((\beta+2N)x+\kappa_1)+(\beta+2N)x^2\pa_x  \nn\\
  &&  +((\beta+2N)xy+\kappa_1y-\kappa_2x)\pa_y+xy\pa_y^2),  \nn\\
R_{+,y} &=& \frac{1}{(\beta+2N)(\beta+2N-1)}((\beta+N-1)((\beta+2N)y+\kappa_2-N)  \nn\\
  &&  +((\beta+2N)xy+(\beta+\kappa_2)x+\kappa_1(1-y))\pa_x+(\beta+2N)y(y-1)\pa_y+x^2\pa_x^2).  \nn
\eea  %
The formulae no longer interchange under the transformation $x\leftrightarrow y, \kappa_1\leftrightarrow \kappa_2$.
These operators satisfy the commutation relations
\bea  %
[L,R_{+,x}] &=& \frac{2x(L-\lambda_N)}{\beta+2N-1} +(\lambda_{N+1}-\lambda_N)R_{+,x}, \nn\\[2mm]
[L,R_{+,y}] &=& \frac{(2y-1)(L-\lambda_N)}{\beta+2N-1} +(\lambda_{N+1}-\lambda_N)R_{+,y}. \nn
\eea  %

Starting from $P_{0,n}$, $R_{+,x}$ is used to build the functions $P_{m,n}, m\geq 1$, with $n$ fixed.  When $n=0$, $R_{+,x}$ reduces to the $1$ dimensional ladder operator
\be\label{r+x2}  %
r_{+,x} = \frac{1}{(\beta+2m)(\beta+2m-1)}((\beta+m-1)((\beta+2m)x+\kappa_1)+(\beta+2m)x^2\pa_x).
\ee  %
On the right edge, $R_{+,y}$ reduces to
$$
r_{+,y} = \frac{1}{(\beta+2m)(\beta+2m-1)}((\beta+n-1)((\beta+2n)y+\kappa_2-n) +(\beta+2n)y(y-1)\pa_y).
$$

\subsection{Krall-Sheffer Metric III}\label{pols3}

\begin{figure}[htb]
\begin{center}
\unitlength=0.3mm
\begin{picture}(160,60)
\put(60,60){\makebox(0,0){$\circ$}}
\put(50,50){\makebox(0,0){$\circ$}}
\put(70,50){\makebox(0,0){$\circ$}}
\put(40,40){\makebox(0,0){$\circ$}}
\put(60,40){\makebox(0,0){{$\circ$}}}
\put(80,40){\makebox(0,0){$\circ$}}
\put(30,30){\makebox(0,0){$\circ$}}
\put(50,30){\makebox(0,0){$\circ$}}
\put(70,30){\makebox(0,0){$\circ$}}
\put(90,30){\makebox(0,0){$\circ$}}
\put(20,20){\makebox(0,0){$\bullet$}}
\put(40,20){\makebox(0,0){$\bullet$}}
\put(60,20){\makebox(0,0){$\bullet$}}
\put(80,20){\makebox(0,0){$\bullet$}}
\put(100,20){\makebox(0,0){$\bullet$}}
\put(10,10){\makebox(0,0){{$\circ$}}}
\put(30,10){\makebox(0,0){{$\circ$}}}
\put(50,10){\makebox(0,0){{$\bullet$}}}
\put(70,10){\makebox(0,0){{$\bullet$}}}
\put(90,10){\makebox(0,0){{$\bullet$}}}
\put(110,10){\makebox(0,0){{$\circ$}}}
\put(0,0){\makebox(0,0){$\circ$}}
\put(20,0){\makebox(0,0){{$\circ$}}}
\put(40,0){\makebox(0,0){$\circ$}}
\put(60,0){\makebox(0,0){$\circ$}}
\put(80,0){\makebox(0,0){$\bullet$}}
\put(100,0){\makebox(0,0){$\circ$}}
\put(120,0){\makebox(0,0){$\circ$}}
\end{picture}
\end{center}
\caption{The pattern for the difference equations for both $P_{m+1,n}$ and $P_{m,n+1}$ for case III.}\label{diff3fig}
\end{figure}
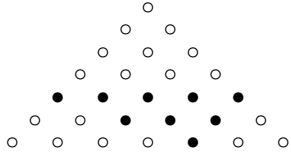

Here we have
\bea  %
L &=& x^2 \pa_x^2+2 x y \pa_x\pa_y+ (y^2+x) \pa_y^2+ (\beta x+\kappa_1)\pa_x+ (\beta y+\kappa_2)\pa_y, \nn\\
 I_1 &=& 2x^2\pa_x\pa_y+xy\pa_y^2+(\kappa_2x-\kappa_1y)\pa_x+(\beta x+\kappa_1)\pa_y,\nn\\
  I_2 &=& x^2\pa_y^2+(\kappa_2x-\kappa_1y)\pa_y. \nn
\eea  %
The polynomial eigenfunctions of $L$ satisfy the $9$ point difference equations:
\bea  %
P_{m+1,n} &=& x P_{m,n}+\frac{1}{\gamma_{2N}\gamma_{2N-2}}\, (\kappa_1(\beta+2n-2) P_{m,n}-2n\kappa_2P_{m+1,n-1}-2n(n-1)P_{m+2,n-2}) \nn\\
&& +\frac{1}{\gamma_{2N-1}\gamma_{2N-2}^2\gamma_{2N-3}}\, (m\kappa_1^2(\beta+m+2n-2) P_{m-1,n} +n\kappa_1\kappa_2(\beta+2n-3) P_{m,n-1}  \nn\\
&& +n(n-1)(\kappa_1(\beta+2n-4)-\kappa_2^2)P_{m+1,n-2} -2n(n-1)(n-2)\kappa_2 P_{m+2,n-3}  \nn\\
&& \qquad\qquad -n(n-1)(n-2)(n-3)P_{m+3,n-4}),  \label{3pmp1n}
\eea  %
and
\bea  %
P_{m,n+1} &=& y P_{m,n}  \nn\\
&&   +\frac{1}{\gamma_{2N}\gamma_{2N-2}}\,
   (\kappa_2(\beta+2m-2)P_{m,n} + 2(n(\beta+2m+n-1) P_{m+1,n-1} -m\kappa_1 P_{m-1,n+1}))  \nn\\
&& +\frac{1}{\gamma_{2N-1}\gamma_{2N-2}^2\gamma_{2N-3}}\, (n(n-1)(n-2)(2\beta+4m+3n-3)P_{m+2,n-3}  \nn\\
&& +n(n-1)(2\beta+6m+4n-5)\kappa_2P_{m+1,n-2} +n((\beta+2m+n-2)(\kappa_2^2-\kappa_1(\beta+3n-3)) \nn\\
&&  +\kappa_1(1-n^2)) P_{m,n-1} +m\kappa_1\kappa_2(\beta+2m-3)P_{m-1,n} -m(m-1)\kappa_1^2P_{m-2,n+1}),  \label{3pmnp1}
\eea  %
where $N=m+n, \gamma_n=\beta+n$. The $9$ points of these difference equations are depicted by the bullet points in Figure \ref{diff3fig}.

Note that in this case, the operator $L$ does not reduce to a 1 dimensional operator on the right edge and, similarly, the difference equations cannot be reduced here.  However on the left edge, the operator $L$ reduces to $L^{(x)}$ of (\ref{LxLy2}) and the difference equation (\ref{3pmp1n}) reduces to (\ref{2pmp10}).  Nevertheless, on the right edge, the difference equation (\ref{3pmnp1}) still makes sense, so can be used to construct the (2 variable) polynomials on the right edge.

Starting from the polynomials $P_{m,0}$ we may use the commuting operators to calculate the remaining polynomials on each level.  By looking at the leading order (monic) part of $I_kP_{m,n}$, we obtain:
\bea  %
I_1P_{m,n}-m\kappa_2P_{m,n} &=& -m\kappa_1P_{m-1,n+1}+n(\beta+2m+n-1)P_{m+1,n-1} ,\nn\\
I_2P_{m,n}+n\kappa_1P_{m,n} &=& n\kappa_2P_{m+1,n-1}+n(n-1)P_{m+2,n-2} ,\nn
\eea  %
which are each {\bf 3-point} differential, recurrence relations on each level.  At the beginning, when $n=0, m=N$, the first gives a 2-point relation, giving $P_{N-1,1}$.  When $m=0, n=N$, we have $I_1P_{0,N}=N(\beta+N-1)P_{1,N-1}$.  Having constructed the polynomials $P_{m,n}, m+n=N$ in this way, we may operate with $I_2$ to move to the left. In this case $I_2P_{N,0}=0$.

Finally, we may construct the operators (\ref{r+}), which are given by
\bea  %
R_{+,x} &=& \frac{1}{\gamma_{2N}\gamma_{2N-1}}\,(\gamma_{N-1}(\gamma_{2N}x+\kappa_1)+\gamma_{2N}x^2\pa_x   +(\gamma_{2N}xy+\kappa_1y-\kappa_2x)\pa_y-x^2\pa_y^2),  \nn\\
R_{+,y} &=& \frac{1}{\gamma_{2N}\gamma_{2N-1}}\,(\gamma_{N-1}\,(\gamma_{2N}y+\kappa_2) +(\gamma_{2N}xy+\kappa_2x-\kappa_1y)\pa_x  \nn\\
  && \hspace{3cm} +(\gamma_{2N}y^2+2\gamma_{N}x+\kappa_1)\pa_y+2x^2\pa_x\pa_y+xy\pa_y^2).  \nn
\eea  %
The formulae no longer interchange under the transformation $x\leftrightarrow y, \kappa_1\leftrightarrow \kappa_2$.
These operators satisfy the commutation relations
\bea  %
[L,R_{+,x}] &=& \frac{2x(L-\lambda_N)}{\gamma_{2N-1}} +(\lambda_{N+1}-\lambda_N)R_{+,x}, \nn\\[2mm]
[L,R_{+,y}] &=& \frac{2y(L-\lambda_N)}{\gamma_{2N-1}} +(\lambda_{N+1}-\lambda_N)R_{+,y}. \nn
\eea  %
On the left edge, $R_{+,x}$ reduces to the $1$ dimensional operator (\ref{r+x2}), which generates the polynomials $P_{m,0}$.  On the right edge, $R_{+,y}$ {\em does not} reduce to a $1$ dimensional operator, but still generates the polynomials $P_{0,n}$.

\subsection{Krall-Sheffer Metric V}\label{pols5}

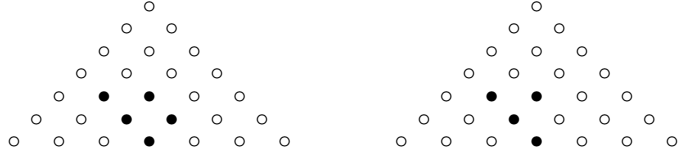
\begin{figure}[htb]
\begin{center}
\subfigure{
\unitlength=0.3mm
\begin{picture}(160,60)
\put(60,60){\makebox(0,0){$\circ$}}
\put(50,50){\makebox(0,0){$\circ$}}
\put(70,50){\makebox(0,0){$\circ$}}
\put(40,40){\makebox(0,0){$\circ$}}
\put(60,40){\makebox(0,0){{$\circ$}}}
\put(80,40){\makebox(0,0){$\circ$}}
\put(30,30){\makebox(0,0){$\circ$}}
\put(50,30){\makebox(0,0){$\circ$}}
\put(70,30){\makebox(0,0){$\circ$}}
\put(90,30){\makebox(0,0){$\circ$}}
\put(20,20){\makebox(0,0){$\circ$}}
\put(40,20){\makebox(0,0){$\bullet$}}
\put(60,20){\makebox(0,0){$\bullet$}}
\put(80,20){\makebox(0,0){$\circ$}}
\put(100,20){\makebox(0,0){$\circ$}}
\put(10,10){\makebox(0,0){{$\circ$}}}
\put(30,10){\makebox(0,0){{$\circ$}}}
\put(50,10){\makebox(0,0){{$\bullet$}}}
\put(70,10){\makebox(0,0){{$\bullet$}}}
\put(90,10){\makebox(0,0){{$\circ$}}}
\put(110,10){\makebox(0,0){{$\circ$}}}
\put(0,0){\makebox(0,0){$\circ$}}
\put(20,0){\makebox(0,0){{$\circ$}}}
\put(40,0){\makebox(0,0){$\circ$}}
\put(60,0){\makebox(0,0){$\bullet$}}
\put(80,0){\makebox(0,0){$\circ$}}
\put(100,0){\makebox(0,0){$\circ$}}
\put(120,0){\makebox(0,0){$\circ$}}
\end{picture}
}
\subfigure{
\unitlength=0.3mm
\begin{picture}(160,60)
\put(60,60){\makebox(0,0){$\circ$}}
\put(50,50){\makebox(0,0){$\circ$}}
\put(70,50){\makebox(0,0){$\circ$}}
\put(40,40){\makebox(0,0){$\circ$}}
\put(60,40){\makebox(0,0){{$\circ$}}}
\put(80,40){\makebox(0,0){$\circ$}}
\put(30,30){\makebox(0,0){$\circ$}}
\put(50,30){\makebox(0,0){$\circ$}}
\put(70,30){\makebox(0,0){$\circ$}}
\put(90,30){\makebox(0,0){$\circ$}}
\put(20,20){\makebox(0,0){$\circ$}}
\put(40,20){\makebox(0,0){$\bullet$}}
\put(60,20){\makebox(0,0){$\bullet$}}
\put(80,20){\makebox(0,0){$\circ$}}
\put(100,20){\makebox(0,0){$\circ$}}
\put(10,10){\makebox(0,0){{$\circ$}}}
\put(30,10){\makebox(0,0){{$\circ$}}}
\put(50,10){\makebox(0,0){{$\bullet$}}}
\put(70,10){\makebox(0,0){{$\circ$}}}
\put(90,10){\makebox(0,0){{$\circ$}}}
\put(110,10){\makebox(0,0){{$\circ$}}}
\put(0,0){\makebox(0,0){$\circ$}}
\put(20,0){\makebox(0,0){{$\circ$}}}
\put(40,0){\makebox(0,0){$\circ$}}
\put(60,0){\makebox(0,0){$\bullet$}}
\put(80,0){\makebox(0,0){$\circ$}}
\put(100,0){\makebox(0,0){$\circ$}}
\put(120,0){\makebox(0,0){$\circ$}}
\end{picture}
}
\end{center}
\caption{The pattern for the difference equations for (a)\, $P_{m+1,n}$, (b)\, $P_{m,n+1}$.}\label{diff5fig}
\end{figure}

Here we have
\bea  %
L &=& 2 x \pa_x\pa_y+ y \pa_y^2+ (\beta x+\kappa_1)\pa_x+ (\beta y+\kappa_2)\pa_y, \nn\\
 I_1 &=& x^2\pa_x^2+(\kappa_2x-\kappa_1y)\pa_x,\nn\\
  I_2 &=& x\pa_y^2+(\beta x+\kappa_1)\pa_y. \nn
\eea  %
The polynomial eigenfunctions of $L$ satisfy the $5$ and $4$ point difference equations:
\be  %
P_{m+1,n} =  \left(x+\frac{\kappa_1}{\beta}\right)P_{m,n}+\frac{2n}{\beta}P_{m+1,n-1}-\frac{n}{\beta^2}((n-1)P_{m+1,n-2}
         +\kappa_1P_{m,n-1}),  \label{5pmp1n}
\ee  %
and
\be  %
P_{m,n+1}=\left(y+\frac{\kappa_2+2(m+n)}{\beta}\right)P_{m,n}-\frac{1}{\beta^2}(n(\kappa_2+2m+n-1)P_{m,n-1}+m\kappa_1P_{m-1,n}).  \label{5pmnp1}
\ee  %
The points of these difference equations are depicted by the bullet points in Figure \ref{diff5fig}.

On the left edge (\ref{5pmp1n}) reduces to
$$
P_{m+1,0} =  \left(x+\frac{\kappa_1}{\beta}\right)P_{m,0}\quad\Rightarrow\quad P_{m,0}= \left(x+\frac{\kappa_1}{\beta}\right)^m.
$$
On the right edge, (\ref{5pmnp1}) reduces to
$$
P_{0,n+1}=\left(y+\frac{\kappa_2+2n}{\beta}\right)P_{0,n}-\frac{1}{\beta^2}(n(\kappa_2+n-1)P_{0,n-1}),
$$
which is the {\em three point recurrence relation} for the Laguerre polynomials.

Notice that the operator $L$ reduces to
$$
(\beta x+\kappa_1)\frac{\pa P_{m,0}}{\pa x}=m\beta P_{m,0},
$$
on the left edge, giving the simple solution above for $P_{m,0}$.  On the right edge, the operator reduces to that of Laguerre.

Starting from the polynomials $P_{m,0}$ we may use the commuting operators to calculate the remaining polynomials on each level.  By looking at the leading order (monic) part of $I_kP_{m,n}$, we obtain:
$$
I_1P_{m,n}-m(\kappa_2+m-1)P_{m,n} = -m\kappa_1P_{m-1,n+1}, \quad I_2P_{m,n}+n\kappa_1P_{m,n} = n\beta P_{m+1,n-1},
$$
with $I_1P_{0,n}=0$ and $I_2P_{m,0}=0$.

Finally, we may construct the operators (\ref{r+}), which are given by
\bea  %
R_{+,x} &=& \frac{1}{\beta^2}\,(x\pa_y^2+(2 \beta x+\kappa_1)\pa_y +\beta (\beta x+\kappa_1)),  \nn\\
R_{+,y} &=& \frac{1}{\beta}\,(x\pa_x + y\pa_y+\beta y+m+n+\kappa_2).  \nn
\eea  %
These operators satisfy the commutation relations
$$
[L,R_{+,x}] = \beta R_{+,x},\quad
[L,R_{+,y}] = \frac{(L-\lambda_N)}{\beta} +\beta R_{+,y}.
$$

On the left edge, $R_{+,x}$ reduces to the {\em multiplicative} operator
$$
r_{+,x} = x+\frac{\kappa_1}{\beta},
$$
giving the above simple formula for $P_{m,0}$.  On the right edge, $R_{+,y}$ reduces to
$$
r_{+,y} = \frac{1}{\beta}\,(y\pa_y+\beta y+n+\kappa_2).
$$

\subsection{Krall-Sheffer Metric VIII}\label{pols8}

\begin{figure}[htb]
\begin{center}
\subfigure{
\unitlength=0.3mm
\begin{picture}(160,60)
\put(60,60){\makebox(0,0){$\circ$}}
\put(50,50){\makebox(0,0){$\circ$}}
\put(70,50){\makebox(0,0){$\circ$}}
\put(40,40){\makebox(0,0){$\circ$}}
\put(60,40){\makebox(0,0){{$\circ$}}}
\put(80,40){\makebox(0,0){$\circ$}}
\put(30,30){\makebox(0,0){$\circ$}}
\put(50,30){\makebox(0,0){$\circ$}}
\put(70,30){\makebox(0,0){$\circ$}}
\put(90,30){\makebox(0,0){$\circ$}}
\put(20,20){\makebox(0,0){$\circ$}}
\put(40,20){\makebox(0,0){$\circ$}}
\put(60,20){\makebox(0,0){$\bullet$}}
\put(80,20){\makebox(0,0){$\bullet$}}
\put(100,20){\makebox(0,0){$\circ$}}
\put(10,10){\makebox(0,0){{$\circ$}}}
\put(30,10){\makebox(0,0){{$\circ$}}}
\put(50,10){\makebox(0,0){{$\circ$}}}
\put(70,10){\makebox(0,0){{$\bullet$}}}
\put(90,10){\makebox(0,0){{$\bullet$}}}
\put(110,10){\makebox(0,0){{$\circ$}}}
\put(0,0){\makebox(0,0){$\circ$}}
\put(20,0){\makebox(0,0){{$\circ$}}}
\put(40,0){\makebox(0,0){$\circ$}}
\put(60,0){\makebox(0,0){$\bullet$}}
\put(80,0){\makebox(0,0){$\circ$}}
\put(100,0){\makebox(0,0){$\circ$}}
\put(120,0){\makebox(0,0){$\circ$}}
\end{picture}
}
\subfigure{
\unitlength=0.3mm
\begin{picture}(160,60)
\put(60,60){\makebox(0,0){$\circ$}}
\put(50,50){\makebox(0,0){$\circ$}}
\put(70,50){\makebox(0,0){$\circ$}}
\put(40,40){\makebox(0,0){$\circ$}}
\put(60,40){\makebox(0,0){{$\circ$}}}
\put(80,40){\makebox(0,0){$\circ$}}
\put(30,30){\makebox(0,0){$\circ$}}
\put(50,30){\makebox(0,0){$\circ$}}
\put(70,30){\makebox(0,0){$\circ$}}
\put(90,30){\makebox(0,0){$\circ$}}
\put(20,20){\makebox(0,0){$\circ$}}
\put(40,20){\makebox(0,0){$\circ$}}
\put(60,20){\makebox(0,0){$\bullet$}}
\put(80,20){\makebox(0,0){$\circ$}}
\put(100,20){\makebox(0,0){$\circ$}}
\put(10,10){\makebox(0,0){{$\circ$}}}
\put(30,10){\makebox(0,0){{$\circ$}}}
\put(50,10){\makebox(0,0){{$\bullet$}}}
\put(70,10){\makebox(0,0){{$\circ$}}}
\put(90,10){\makebox(0,0){{$\circ$}}}
\put(110,10){\makebox(0,0){{$\circ$}}}
\put(0,0){\makebox(0,0){$\circ$}}
\put(20,0){\makebox(0,0){{$\circ$}}}
\put(40,0){\makebox(0,0){$\circ$}}
\put(60,0){\makebox(0,0){$\bullet$}}
\put(80,0){\makebox(0,0){$\circ$}}
\put(100,0){\makebox(0,0){$\circ$}}
\put(120,0){\makebox(0,0){$\circ$}}
\end{picture}
}
\end{center}
\caption{The pattern for the difference equations for (a)\, $P_{m+1,n}$, (b)\, $P_{m,n+1}$.}\label{diff8fig}
\end{figure}
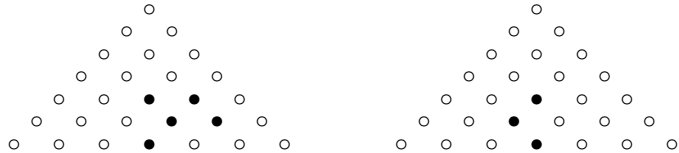

Here we have
\bea  %
L &=& y \pa_x^2+2 \pa_x\pa_y+  (\beta x+\kappa_1)\pa_x+ (\beta y+\kappa_2)\pa_y, \nn\\
 I_1 &=& \pa_x^2+(\beta y+\kappa_2)\pa_x,\nn\\
  I_2 &=& (y^2-x)\pa_x^2+2y\pa_x\pa_y+\pa_y^2+(\kappa_1 y-\kappa_2 x)\pa_x+(\beta x+\kappa_1)\pa_y. \nn
\eea  %
The polynomial eigenfunctions of $L$ satisfy the $5$ and $3$ point difference equations:
\be  %
P_{m+1,n} =  \left(x+\frac{\kappa_1}{\beta}\right)P_{m,n}+\frac{1}{\beta}\, (2m P_{m-1,n+1}+ n P_{m,n-1})
    -\frac{m \kappa_2}{\beta^2}\, P_{m-1,n},  \label{8pmp1n}
\ee  %
and
\be  %
P_{m,n+1}=\left(y+\frac{\kappa_2}{\beta}\right)P_{m,n}+\frac{m}{\beta} P_{m-1,n}.  \label{8pmnp1}
\ee  %
The points of these difference equations are depicted by the bullet points in Figure \ref{diff8fig}.

On the right edge, (\ref{8pmnp1}) reduces to
$$
P_{0,n+1}=\left(y+\frac{\kappa_2}{\beta}\right)P_{0,n}\quad\Rightarrow\quad P_{0,n}=\left(y+\frac{\kappa_2}{\beta}\right)^n.
$$
On the left edge (\ref{8pmp1n}) reduces to
$$
P_{m+1,0} =   \left(x+\frac{\kappa_1}{\beta}\right)P_{m,0}+\frac{2m}{\beta}\, P_{m-1,1}-\frac{m \kappa_2}{\beta^2}\, P_{m-1,0},
$$
which is no longer a $3$ point recursion on the edge, since the left edge polynomials are not functions of just $x$.

Starting from the polynomials $P_{0,n}$ we may use the commuting operators to calculate the remaining polynomials on each level.  By looking at the leading order (monic) part of $I_kP_{m,n}$, we obtain:
\bea  %
I_1P_{m,n} &=& m\beta P_{m-1,n+1}, \nn\\
I_2P_{m,n}+m\kappa_2P_{m,n} &=& n\beta P_{m+1,n-1}+m\kappa_1 P_{m-1,n+1}+m(m-1) P_{m-2,n+2},\nn
\eea  %
with $I_1P_{0,n}=0$.

Finally, we may construct the operators (\ref{r+}), which are given by
\bea  %
R_{+,x} &=& x+\frac{\kappa_1}{\beta} + \frac{1}{\beta}\, \pa_y+ \frac{1}{\beta^2}\, ((2 \beta y+\kappa_2)\pa_x+\pa_x^2),  \nn\\
R_{+,y} &=& y+\frac{\kappa_2}{\beta} + \frac{1}{\beta}\,\pa_x.  \nn
\eea  %
These operators satisfy the commutation relations
$$
[L,R_{+,x}] = \beta R_{+,x},\quad  [L,R_{+,y}] = \beta R_{+,y}.
$$

On the right edge, $R_{+,y}$ reduces to the {\em multiplicative} operator
$$
r_{+,y} = y+\frac{\kappa_2}{\beta},
$$
giving the above simple formula for $P_{0,n}$.  On the left edge, $R_{+,x}$ does not reduce to an operator in just $x$, but still generates the functions $P_{m,0}$ from $P_{0,0}=1$.

\subsection{Krall-Sheffer Metric IX}\label{pols9}

In this case it is convenient to consider the 4 operators, which commute with $L$:
\bea  %
&& L= (x^2-1) \pa_x^2+2 x y \pa_x\pa_y+ (y^2-1) \pa_y^2+ \beta x\pa_x+ \beta y\pa_y,  \nn\\
&& I_1=(1-x^2-y^2)\pa_x^2+(1-\beta) x \pa_x,\quad  I_2=(1-x^2-y^2)\pa_y^2+(1-\beta) y \pa_y,  \label{case9-4ops} \\
&&  I_3= x \pa_y - y \pa_x,\quad  I_4= 2(1-x^2-y^2)\pa_x\pa_y+(1-\beta) (x \pa_y + y \pa_x).  \nn
\eea  %
Under the commutator, these form a closed, quadratic algebra (see \cite{scott}), which we don't need here. They are not, of course, independent.
They satisfy the quadratic relations:
$$
I_1+I_2+I_3^2+L=0,\quad 2(I_1I_2+I_2I_1)-(\beta^2-4\beta-1)(I_1+I_2)=(\beta-1)(\beta-5)L+I_4^2.
$$
These also satisfy
$$
\begin{array}{l}
I_1P_{m,n} = (2-\beta-m)m P_{m,n}-m(m-1)P_{m-2,n+2}, \\[1mm]
I_2P_{m,n} = (2-\beta-n)n P_{m,n}-n(n-1)P_{m+2,n-2},\\[1mm]
I_3P_{m,n} = n P_{m+1,n-1}-m P_{m-1,n+1}, \\[1mm]
I_4P_{m,n} = (1-\beta-2m)n P_{m+1,n-1}+(1-\beta-2n) m P_{m-1,n+1},
\end{array}
$$
so we use $I_3$ and $I_4$ (relating ``nearest neighbours'') as our horizontal ladders.

\begin{figure}[htb]
\begin{center}
\subfigure{
\unitlength=0.3mm
\begin{picture}(160,60)
\put(60,60){\makebox(0,0){$\circ$}}
\put(50,50){\makebox(0,0){$\circ$}}
\put(70,50){\makebox(0,0){$\circ$}}
\put(40,40){\makebox(0,0){$\circ$}}
\put(60,40){\makebox(0,0){{$\circ$}}}
\put(80,40){\makebox(0,0){$\circ$}}
\put(30,30){\makebox(0,0){$\circ$}}
\put(50,30){\makebox(0,0){$\circ$}}
\put(70,30){\makebox(0,0){$\circ$}}
\put(90,30){\makebox(0,0){$\circ$}}
\put(20,20){\makebox(0,0){$\circ$}}
\put(40,20){\makebox(0,0){$\bullet$}}
\put(60,20){\makebox(0,0){$\circ$}}
\put(80,20){\makebox(0,0){$\bullet$}}
\put(100,20){\makebox(0,0){$\circ$}}
\put(10,10){\makebox(0,0){{$\circ$}}}
\put(30,10){\makebox(0,0){{$\circ$}}}
\put(50,10){\makebox(0,0){{$\circ$}}}
\put(70,10){\makebox(0,0){{$\bullet$}}}
\put(90,10){\makebox(0,0){{$\circ$}}}
\put(110,10){\makebox(0,0){{$\circ$}}}
\put(0,0){\makebox(0,0){$\circ$}}
\put(20,0){\makebox(0,0){{$\circ$}}}
\put(40,0){\makebox(0,0){$\circ$}}
\put(60,0){\makebox(0,0){$\bullet$}}
\put(80,0){\makebox(0,0){$\circ$}}
\put(100,0){\makebox(0,0){$\circ$}}
\put(120,0){\makebox(0,0){$\circ$}}
\end{picture}
}
\subfigure{
\unitlength=0.3mm
\begin{picture}(160,60)
\put(60,60){\makebox(0,0){$\circ$}}
\put(50,50){\makebox(0,0){$\circ$}}
\put(70,50){\makebox(0,0){$\circ$}}
\put(40,40){\makebox(0,0){$\circ$}}
\put(60,40){\makebox(0,0){{$\circ$}}}
\put(80,40){\makebox(0,0){$\circ$}}
\put(30,30){\makebox(0,0){$\circ$}}
\put(50,30){\makebox(0,0){$\circ$}}
\put(70,30){\makebox(0,0){$\circ$}}
\put(90,30){\makebox(0,0){$\circ$}}
\put(20,20){\makebox(0,0){$\circ$}}
\put(40,20){\makebox(0,0){$\bullet$}}
\put(60,20){\makebox(0,0){$\circ$}}
\put(80,20){\makebox(0,0){$\bullet$}}
\put(100,20){\makebox(0,0){$\circ$}}
\put(10,10){\makebox(0,0){{$\circ$}}}
\put(30,10){\makebox(0,0){{$\circ$}}}
\put(50,10){\makebox(0,0){{$\bullet$}}}
\put(70,10){\makebox(0,0){{$\circ$}}}
\put(90,10){\makebox(0,0){{$\circ$}}}
\put(110,10){\makebox(0,0){{$\circ$}}}
\put(0,0){\makebox(0,0){$\circ$}}
\put(20,0){\makebox(0,0){{$\circ$}}}
\put(40,0){\makebox(0,0){$\circ$}}
\put(60,0){\makebox(0,0){$\bullet$}}
\put(80,0){\makebox(0,0){$\circ$}}
\put(100,0){\makebox(0,0){$\circ$}}
\put(120,0){\makebox(0,0){$\circ$}}
\end{picture}
}
\end{center}
\caption{The pattern for the difference equations for (a)\, $P_{m+1,n}$, (b)\, $P_{m,n+1}$.}\label{diff9fig}
\end{figure}
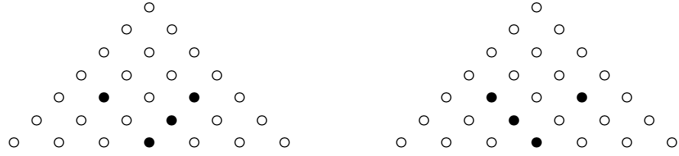

The polynomial eigenfunctions of $L$ satisfy the $4$ point difference equations:
$$
P_{m+1,n} =  x P_{m,n}+\frac{1}{\gamma_{2N-1}\gamma_{2N-3}}\, (n(n-1) P_{m+1,n-2}-m(\beta+m+2n-2) P_{m-1,n}),  \label{9pmp1n}
$$
and
\be  %
P_{m,n+1}= y P_{m,n}+\frac{1}{\gamma_{2N-1}\gamma_{2N-3}}\, (m(m-1) P_{m-2,n+1}-n(\beta+2m+n-2) P_{m,n-1}),  \label{9pmnp1}
\ee  %
where $N=m+n, \gamma_n=\beta+n$.
The points of these difference equations are depicted by the bullet points in Figure \ref{diff9fig}.  On the left edge, the first reduces to
$$
P_{m+1,0} =  x P_{m,0}-\frac{1}{\gamma_{2m-1}\gamma_{2m-3}}\, m(\beta+m-2) P_{m-1,0},
$$
which is the 3 point recurrence relation for Gegenbauer's equation
$$
(1-x^2)\frac{d^2P_{m,0}}{dx^2}-\beta x \frac{dP_{m,0}}{dx}=m(m+\beta-1)P_{m,0}.
$$
The second relation (\ref{9pmnp1}) similarly reduces on the right edge.

Finally, we may construct the operators (\ref{r+}), which are given by
\bea  %
R_{+,x} &=& \frac{1}{\gamma_{2N-1}}\,(x y\pa_y+(x^2-1)\pa_x + (\beta +N-1)x),  \nn\\
R_{+,y} &=& \frac{1}{\gamma_{2N-1}}\,(x y\pa_x+(y^2-1)\pa_y + (\beta +N-1)y).  \nn
\eea  %
These operators satisfy the commutation relations
\bea  %
[L,R_{+,x}] &=& \frac{2x(L-\lambda_N)}{\gamma_{2N-1}} +(\lambda_{N+1}-\lambda_N)R_{+,x}, \nn\\[2mm]
[L,R_{+,y}] &=& \frac{2y(L-\lambda_N)}{\gamma_{2N-1}} +(\lambda_{N+1}-\lambda_N)R_{+,y}. \nn
\eea  %

\section{Generating Functions}\label{genfuns}

Generating functions are known for the special cases IV, VI and VII, since these are just products of the known classical generating functions.
The generating function for Case IX is just a direct generalisation of that for Gegenbauer polynomials (in 1 dimension).  A multi-dimensional generalisation can be found in \cite{01-11}.  We believe that the generating functions given below for Cases V and VIII are new.

Generating functions are important for a number of reasons and can be considered as fundamental.  Suppose a given family of polynomials $P_{m,n}$ are coefficients in a generating function expansion
$$
G(x,y,s,t)=\sum_{m,n=0}^\infty A_{m,n} P_{m,n} \frac{s^m t^n}{m!n!},
$$
where the additional coefficients $A_{m,n}$ are included to allow our polynomials to be {\em monic}.
Then, by considering $\frac{\pa G}{\pa s}$ and $\frac{\pa G}{\pa t}$, we can construct explicit forms of difference equations, relating polynomials of degrees $m+n+1, m+n$ and $m+n-1$.
By considering $\frac{\pa G}{\pa x}$ and $\frac{\pa G}{\pa y}$, we can construct explicit forms of differential ladder operators.  Manipulating these, we can construct the eigenvalue problem for the polynomials, thus proving that they really are eigenfunctions of a particular differential operator.

\subsection{Cases I and IX: Generalised Gegenbauer Polynomials}

Since the operator $L$ of (\ref{case9-4ops}), for Case IX, is symmetric in $x$ and $y$ it is easy to generalise the standard Gegenbauer generating function to
$$
G(x,y,s,t)=(1-2sx-2ty+s^2+t^2)^{(1-\beta)/2}=
  \sum_{m,n=0}^\infty \frac{2^{m+n} \Gamma\left(\frac{\beta+2(m+n)-1}{2}\right) P_{m,n} s^m t^n}{\Gamma\left(\frac{\beta-1}{2}\right)m!n!}.
$$
Notice that this generating function has a pair of symmetries $(s,x)\mapsto (-s,-x)$ and $(t,y)\mapsto (-t,-y)$, which must be reflected in each term $P_{m,n}s^mt^n$. When $m$ is even (odd), $P_{m,n}$ is an even (odd) function of $x$ (and similarly for $n$ and $y$).  This means there will be an even polynomial of $x$, $\rho_{2m,n}$, such that $P_{2m+1,n}=x\rho_{2m,n}$ (with a similar statement for $P_{m,2n+1}$).  Indeed, the first few polynomials are
\bea  %
&& P_{0,0}=1, \;\; P_{1,0}=x,\;\; P_{0,1}=y,\;\; P_{2,0}=x^2-\frac{1}{1+\beta},\;\; P_{1,1}=x y,\;\; P_{0,2}=y^2-\frac{1}{1+\beta},\nn\\
&&  P_{3,0}=x\left(x^2-\frac{3}{3+\beta}\right),\;\; P_{2,1}=y\left(x^2-\frac{1}{3+\beta}\right),\nn\\
&&  P_{1,2}=x\left(y^2-\frac{1}{3+\beta}\right),\;\; P_{0,3}=y\left(y^2-\frac{3}{3+\beta}\right).  \nn
\eea  %
In Figure \ref{pols9fig} below, polynomials which are even in both $x$ {\bf and} $y$ are depicted by black dots.  They occur only when $N=m+n$ is even.
\begin{figure}[htb]
\begin{center}
\unitlength=0.3mm
\begin{picture}(160,60)
\put(60,60){\makebox(0,0){$\bullet$}}
\put(50,50){\makebox(0,0){$\circ$}}
\put(70,50){\makebox(0,0){$\circ$}}
\put(40,40){\makebox(0,0){$\bullet$}}
\put(60,40){\makebox(0,0){{$\circ$}}}
\put(80,40){\makebox(0,0){$\bullet$}}
\put(30,30){\makebox(0,0){$\circ$}}
\put(50,30){\makebox(0,0){$\circ$}}
\put(70,30){\makebox(0,0){$\circ$}}
\put(90,30){\makebox(0,0){$\circ$}}
\put(20,20){\makebox(0,0){$\bullet$}}
\put(40,20){\makebox(0,0){$\circ$}}
\put(60,20){\makebox(0,0){$\bullet$}}
\put(80,20){\makebox(0,0){$\circ$}}
\put(100,20){\makebox(0,0){$\bullet$}}
\put(10,10){\makebox(0,0){{$\circ$}}}
\put(30,10){\makebox(0,0){{$\circ$}}}
\put(50,10){\makebox(0,0){{$\circ$}}}
\put(70,10){\makebox(0,0){{$\circ$}}}
\put(90,10){\makebox(0,0){{$\circ$}}}
\put(110,10){\makebox(0,0){{$\circ$}}}
\put(0,0){\makebox(0,0){$\bullet$}}
\put(20,0){\makebox(0,0){{$\circ$}}}
\put(40,0){\makebox(0,0){$\bullet$}}
\put(60,0){\makebox(0,0){$\circ$}}
\put(80,0){\makebox(0,0){$\bullet$}}
\put(100,0){\makebox(0,0){$\circ$}}
\put(120,0){\makebox(0,0){$\bullet$}}
\end{picture}
\end{center}
\caption{The pattern of polynomials for case IX. The black dots are when {\bf both} $m$ and $n$ are even.}\label{pols9fig}
\end{figure}
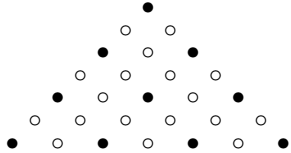
This configuration is unimportant in the context of case IX, but when the transformation $(x_9,y_9)=(\sqrt{x_1},\sqrt{y_1})$ (see Section \ref{i-ix}) is used, the black dots represent the polynomial eigenfunctions in Case I (with the restriction $\kappa_1=\kappa_2=-\frac{1}{2}$).  Generally, $P_{m,n}$ (of Case IX) maps to $P_{\frac{m}{2},\frac{n}{2}}$ (of Case I), giving us ``half-integer'' indices, corresponding to non-polynomial eigenfunctions.  For example, $P_{2,1}\mapsto P_{1,\frac{1}{2}}=\sqrt{y}\left(x-\frac{1}{2(\beta+1)}\right)$.

In Section \ref{pols9} we chose to use $I_3, I_4$ to take us right and left on horizontal levels, since they just related ``nearest neighbours''.  The operators $I_1, I_2$ jump two steps at a time, connecting (in particular) the black dots on the horizontal levels.  These transform to the $I_1,I_2$ given in Section \ref{pols1} (with $\kappa_1=\kappa_2=-\frac{1}{2}$ and the replacement $\beta_9\mapsto 2\beta_1-1$).  With these values of $\kappa_i$, we have
$$
I_1P_{m,n}+m\left(\beta+m-\frac{3}{2}\right)P_{m,n} = m\left(\frac{1}{2}-m\right)P_{m-1,n+1},
$$
so, when $m=\frac{1}{2}$, the right hand side vanishes and $P_{\frac{1}{2},n}$ is just an eigenfunction of $I_1$.  This means that starting on one of the half-integer functions, $I_1$ moves us to the right until we reach the last one on the given level.  The operator $I_2$ acts in a similar way.  The operators $I_3, I_4$ can also be transformed to the reduced Case I, but these operators no longer have polynomial coefficients.

\subsection{Cases V and VIII}

For Case V, the generating function is
$$
G(x,y,s,t)=\exp\left(\frac{\beta^2 s x+(\kappa_1s+\beta t y)(\beta-t)}{(\beta-t)^2}\right)\,\left(1-\frac{t}{\beta}\right)^{-\kappa_2}
= \sum_{m,n=0}^\infty P_{m,n}(x,y) \frac{s^m t^n}{m! n!}.
$$
When $t=0$ this reduces to $G_l(x,s)=\exp\left(s\left(x+\frac{\kappa_1}{\beta}\right)\right)$ and generates the polynomials $P_{m,0}=\left(x+\frac{\kappa_1}{\beta}\right)^m$, located on the left edge.  With $s=0$, the generating function reduces to
$$
G_r(y,t)=\exp\left(\frac{\beta t y}{(\beta-t)}\right)\,\left(1-\frac{t}{\beta}\right)^{-\kappa_2}
$$
and generates Laguerre polynomials $P_{0,n}$ on the right edge.  In the interior of the triangle the polynomials depend on both $x$ and $y$ in a non-trivial way.  For example
$$
P_{1,1}=\left(x+\frac{\kappa_1}{\beta}\right)\left(y+\frac{\kappa_2}{\beta}\right)+\frac{2}{\beta}\left(x+\frac{\kappa_1}{2\beta}\right).
$$

By differentiating $G(x,y,s,t)$ with respect to $s$ we find
$$
(\beta-t)^2\frac{\pa G}{\pa s}=(\beta^2 x+\kappa_1(\beta-t))G,
$$
which leads to the {\em 3 level recurrence relation}
$$
P_{m+1,n}=\left(x+\frac{\kappa_1}{\beta}\right)P_{m,n}+\frac{2n}{\beta}P_{m+1,n-1}-\frac{n}{\beta^2}((n-1)P_{m+1,n-2}+\kappa_1P_{m,n-1}),
$$
which we already saw in Section \ref{pols5}.  When $n=0$, this reduces to $P_{m+1,0}=\left(x+\frac{\kappa_1}{\beta}\right)P_{m,0}$, which gives the above $P_{m,0}$.

Differentiating $G(x,y,s,t)$ with respect to $t$ leads to a more complicated formula, but can be combined with the $s-$derivative to give
$$
(\beta-t)^2\frac{\pa G}{\pa t}-2s(\beta-t)\frac{\pa G}{\pa s}=(\beta^2 y+\kappa_2(\beta-t)-\kappa_1s)G,
$$
which leads to the other {\em 3 level recurrence relation}
$$
P_{m,n+1}=\left(y+\frac{\kappa_2+2(m+n)}{\beta}\right)P_{m,n}-\frac{1}{\beta^2}(n(\kappa_2+2m+n-1)P_{m,n-1}+m\kappa_1P_{m-1,n}),
$$
also given in Section \ref{pols5}.

For Case VIII, the generating function is
\bea  %
G(x,y,s,t) &=& \exp\left(\frac{4s^3+3(2\beta y+\kappa_2)s^2+6\beta s t+6\beta ((\beta x+\kappa_1)s+(\beta y+\kappa_2)t)}{6\beta^2}\right)\nn\\
&=& \sum_{m,n=0}^\infty P_{m,n}(x,y) \frac{s^m t^n}{m! n!}.\nn
\eea  %
Similar manipulations as described above lead to the difference equations (\ref{8pmp1n}) and (\ref{8pmnp1}).

The transformation between the Laplace-Beltrami operators of Cases V and VIII (see Section \ref{v-viii}) do not extend to the Krall Sheffer operators, so the polynomials generated for either case cannot be transformed to eigenfunctions of the other.

\section{Conclusions}

In this paper we presented a number of recursive algebraic structures which allow the explicit construction of polynomial eigenfunctions of the Krall-Sheffer differential operators.  By arranging these polynomials in the triangular array of Figure \ref{trilatt}, we showed that the $2-$dimensional Krall-Sheffer eigenvalue problem reduced to a standard $1-$dimensional problem on at least one of the edges, where we could use the standard {\em $3-$point recurrence relation} to build polynomial eigenfunctions of one variable.  We then used the commuting operators (coming from super-integrability) to build the $2-$dimensional polynomials on the horizontal level of the triangle.

We then gave {\em explicit formulae} for the {\em $3-$level recurrence relations} for each of the (non-trivial) Krall-Sheffer cases.  These reduce to the corresponding $3-$point relations on appropriate edges.  The $3-$level recurrence relations give a very effective way of building the polynomials. We also gave {\em explicit formulae} for the differential raising operators, which shift us to the next eigenvalue in directions parallel to the edges of the triangular array.  Lowering operators also exist (see \cite{scott}) and sometimes have simple formulae.  However, for some of the cases, these operators are very complicated, so we didn't include any of them here.

Finally, we discussed the existence of generating functions for the Krall-Sheffer polynomials.  We presented new generating functions for cases V and VIII.  It was shown how the well known generalised Gegenbauer generating function of Case IX can be used to build eigenfunctions of Case I with special parameter values.  It would be interesting to generalise this to a generating function for the {\em full} Case I.  Generating functions for Cases II and III are also currently unknown.

The discussion of Sections \ref{geom} and \ref{super} give us a direction to follow if we wish to build Krall-Sheffer like operators in higher dimensions.  Some calculations in this direction can be found in \cite{scott}, but it is an important subject for future research.

\subsubsection*{Acknowledgments:}
The authors would like to thank Y. Xu and A.V. Turbiner for bringing the papers \cite{95-6} and \cite{09-5} to their attention.


\begin{thebibliography}{10}

\bibitem{95-6}
H.~Berens, H.~J. Schmid, and Y.~Xu.
\newblock On two-dimensional definite orthogonal systems and a lower bound for
  the number of nodes of associated cubature formulae.
\newblock {\em SIAM J. Math. Anal.}, 26:468--87, 1995.

\bibitem{11-1}
T.~S. Chihara.
\newblock {\em An Introduction to Orthogonal Polynomials}.
\newblock Dover, New York, 2011.

\bibitem{06-4}
A.~M. Delgado, J.~S. Geronimo, P~Iliev, and F.~Marcellan.
\newblock Two variable orthogonal polynomials and structured matrices.
\newblock {\em SIAM J. Matrix Analysis and Applications}, 28:118--47, 2006.

\bibitem{01-11}
C.~F. Dunkl and Y.~Xu.
\newblock {\em Orthogonal Polynomials of Several Variables}.
\newblock CUP, Cambridge, 2001.

\bibitem{f07-1}
A.P. Fordy.
\newblock Quantum super-integrable systems as exactly solvable models.
\newblock {\em SIGMA}, 3:025, 10 pages, 2007.
\newblock http://dx.doi.org/10.3842/SIGMA.2007.025.

\bibitem{74-7}
R.~Gilmore.
\newblock {\em Lie Groups, Lie Algebras and Some of Their Applications}.
\newblock Wiley, New York, 1974.

\bibitem{01-1}
J.~Harnad, L.~Vinet, O.~Yermolayeva, and A.~Zhedanov.
\newblock Two-dimensional {Krall-Sheffer} polynomials and integrable systems.
\newblock {\em J.Phys.A}, 34:10619--25, 2001.

\bibitem{99-11}
E.G. Kalnins, Jr. W.~Miller, Ye.M. Hakobyan, and G.S. Pogosyan.
\newblock Superintegrability on the two-dimensional hyperboloid. {II}.
\newblock {\em J.Math.Phys.}, 40:2291--2306, 1999.

\bibitem{67-2}
H.L. Krall and I.M. Sheffer.
\newblock Orthogonal polynomials in two variables.
\newblock {\em Ann.Mat.Pura Appl. ser. 4}, 76:325--76, 1967.

\bibitem{76-8}
L.D. Landau and E.M. Lifshitz.
\newblock {\em Course of Theoretical Physics Volume 1: Mechanics}.
\newblock Pergamon, Oxford, 1976.

\bibitem{scott}
M.J. Scott.
\newblock {\em Classical and quantum integrable systems on manifolds with
  symmetry}.
\newblock PhD thesis, University of Leeds, 2010.

\bibitem{99-13}
P.K. Suetin.
\newblock {\em Orthogonal Polynomials in Two Variables}.
\newblock Gordon and Breach, Amsterdam, 1999.

\bibitem{09-5}
F.~Tremblay, A.~V. Turbiner, and P.~Winternitz.
\newblock An infinite family of solvable and integrable quantum systems on a
  plane.
\newblock {\em J Phys A}, 42:242001, 2009.

\bibitem{03-4}
L.~Vinet and A.~Zhedanov.
\newblock Two-dimensional {Krall-Sheffer} polynomials and quantum systems on
  spaces of constant curvature.
\newblock {\em Letts.Math.Phys.}, 65:83--94, 2003.

\end{thebibliography}

\end{document}